\documentclass[3p]{elsarticle}
\UseRawInputEncoding
\usepackage{lineno,hyperref}
\usepackage{graphicx}
\usepackage{amsmath}
\usepackage{color}
\usepackage[normalem]{ulem}
\usepackage{mathtools}
\usepackage{rotating}
\usepackage{lscape}
\usepackage{booktabs}
\usepackage{subfigure}
\usepackage{placeins}
\usepackage{array}
\usepackage{tabularx}

\journal{arXiv}
\begin{document}

\begin{frontmatter}

\title{Broken Symmetry, Conservation Law, and Scaling in Accumulated Stock Returns -- a Modified Jones-Faddy Skew t-Distribution Perspective} 

\author[mymainaddress]{Arshia Ghasemi}
\author[mymainaddress]{Siqi Shao}
%\author[mymainaddress]{Hamed Farahani}
\author[mymainaddress]{R. A. Serota\fnref{myfootnote}}
\fntext[myfootnote]{serota@ucmail.uc.edu}

\address[mymainaddress]{Department of Physics, University of Cincinnati, Cincinnati, Ohio 45221-0011}

\begin{abstract}
We analyze historic S\&P500 multi-day returns: from daily returns to those accumulated over up to ten days. Despite symmetry breaking between gains and losses in the distribution of returns, resulting in its positive mean and negative skew, realized variance (volatility squared) exhibits remarkably good linear dependence on the number of days of accumulation. Mean of the distribution also shows near perfect linear dependence as well. We analyze this phenomenon both analytically and numerically using a modified Jones-Faddy skew t-distribution.
\end{abstract}

\begin{keyword}
Student's  t-Distribution \sep Jones-Faddy Skew t-Distribution \sep Accumulated Stock Returns Gains and Losses \sep Stochastic Volatility \sep Power-Law Tails
\end{keyword}

\end{frontmatter}

\section{Introduction}

\begin{figure}[htbp]
    \centering
    \begin{tabular}{cc}
\includegraphics[width=0.49\linewidth]{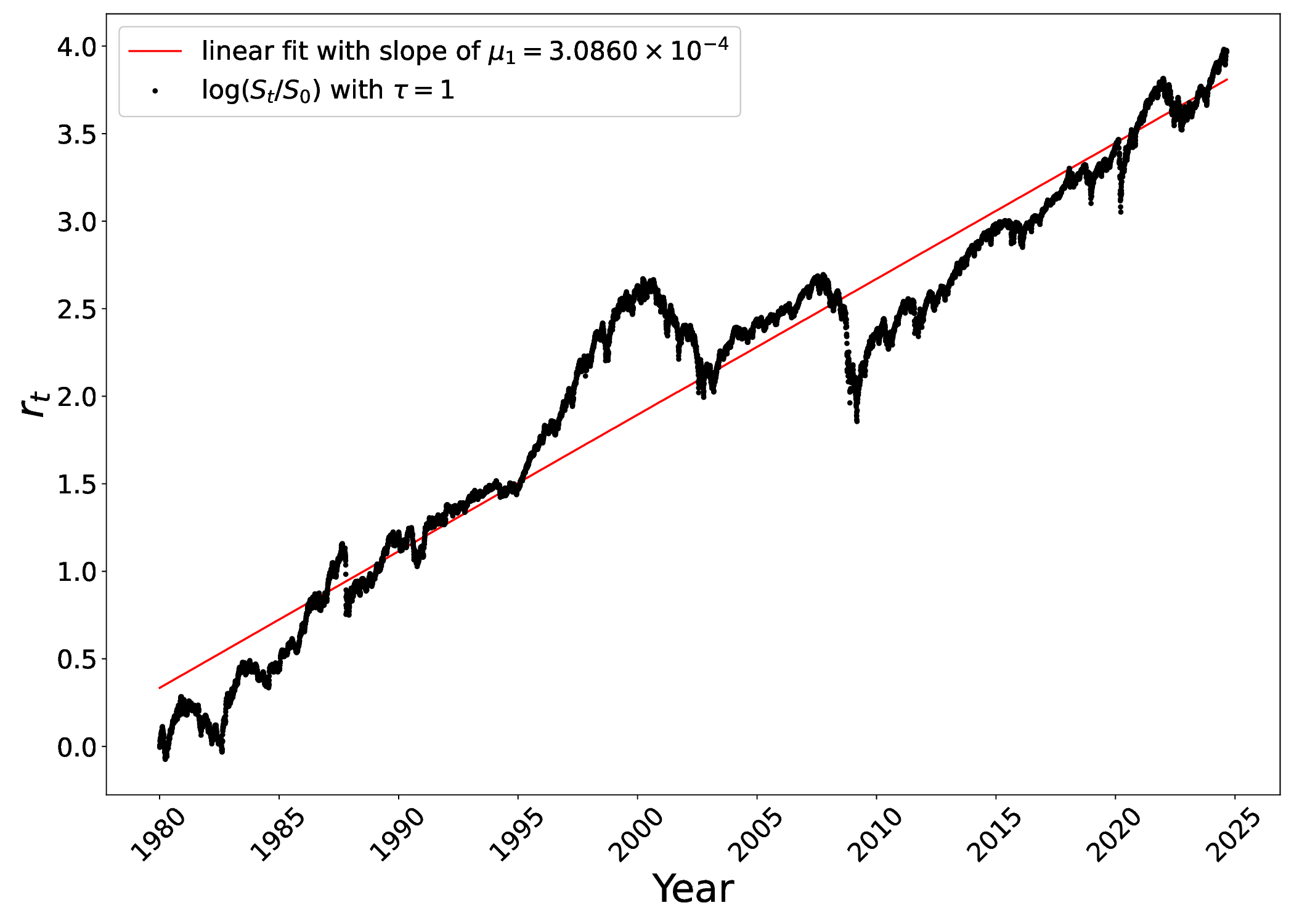} &
 \includegraphics[width=0.49\linewidth]{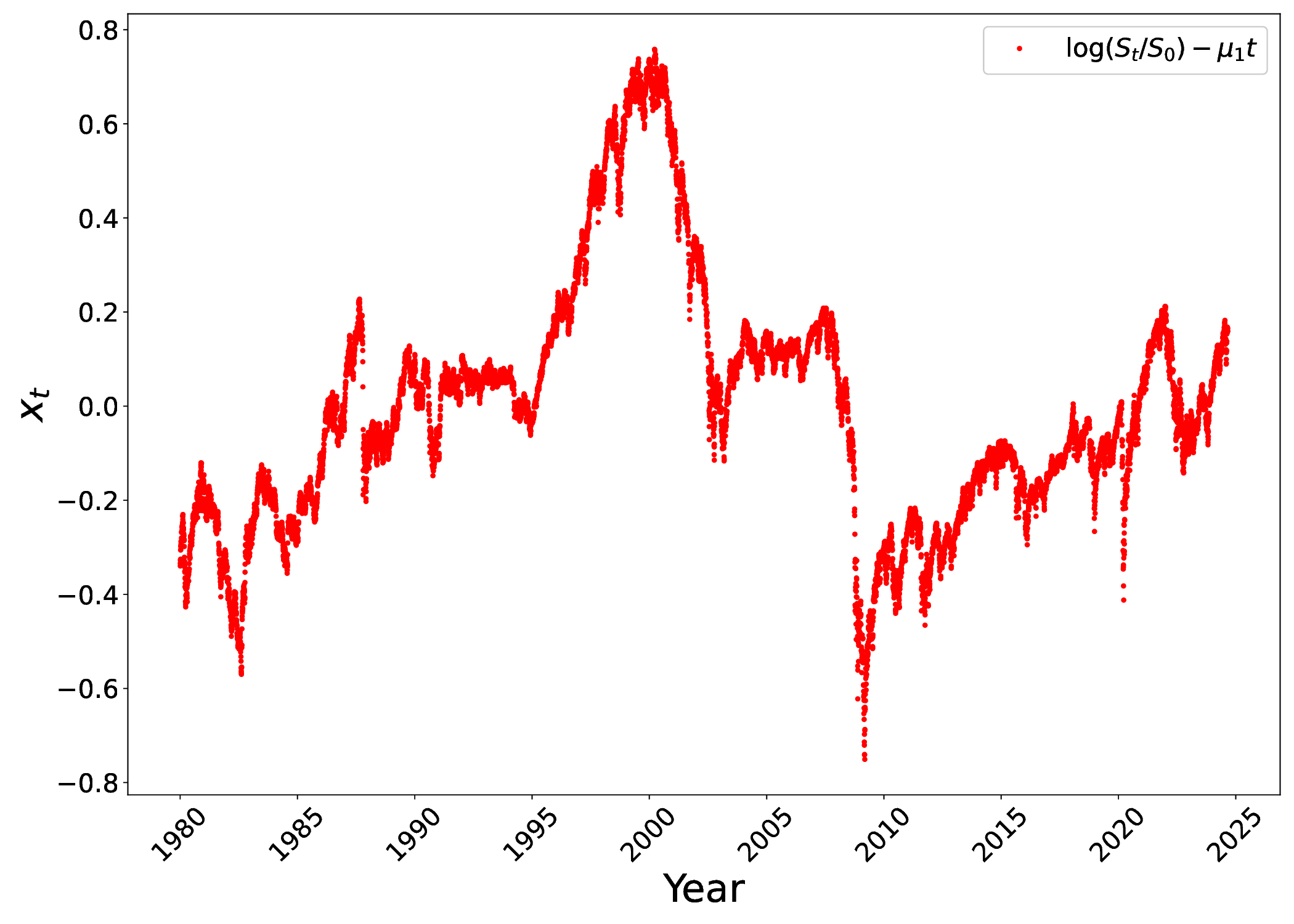} \\
   \end{tabular}
    \caption{S\&P500. Left: $r_t=\log(S_t/S_0)$, $S_t$ is price on day $t$, $t$ changes in daily increments ($\tau =1$ in text); Right:S\&P500; $x_t = r_t - \mu_1 t$ where index in $\mu_1$ reflects daily increments of $t$ ($\tau =1$ in text). From \cite {farahani2025asymmetry}.}
    \label{rtxt}
\end{figure}

\begin{figure}[!htbp]
\centering
\begin{tabular}{cc}
\includegraphics[width = 0.49 \textwidth]{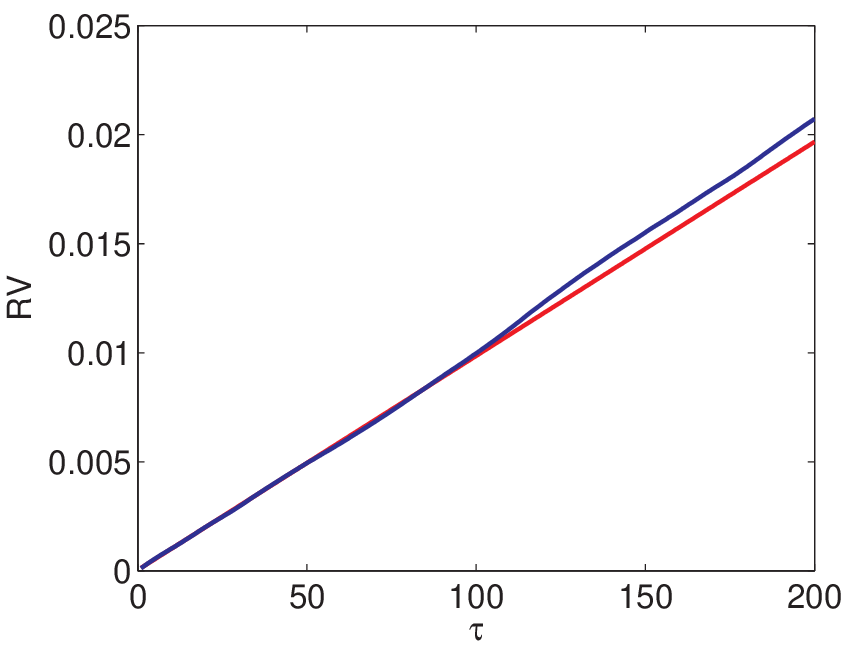} &
\includegraphics[width = 0.49 \textwidth]{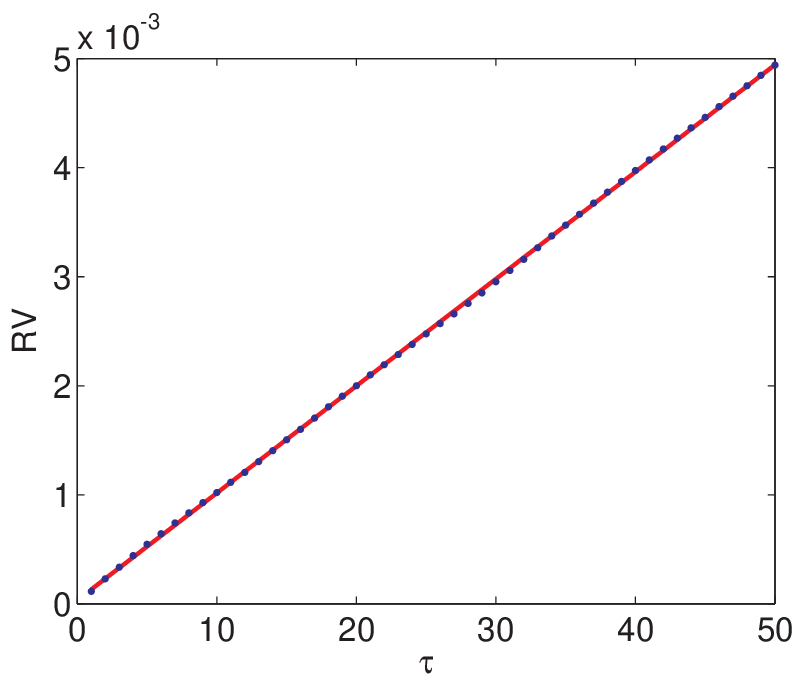}  \\
\end{tabular}
\caption{Realized variance (RV) of S\&P500 as a function of the number of days 
$\tau$ over which the returns are calculated. The best straight line fit is $f(\tau) = 1.062 \times 10^{-4} \tau - 3.328 \times 
10^{-5}$. From \cite{liu2019distributions}.}
\label{RV}
\end{figure}

The left-hand side plot of Fig. \ref{rtxt} shows linear fit, $\mu_1 t$, of $r_t = \log(S_t/S_0)$ for S\&P500, where $S_t$ is price on day $t$.\footnote{We used daily S\&P500 returns, hence index ``1" in $\mu_1$. However the magnitude of the slope depends only slightly on the number of days used as time steps \cite{farahani2025asymmetry}.} This upward trend corresponds to roughly $12\%$ annual growth and is a beloved number cited by brokers and financial advisors (maybe $10\%$ on the account of inflation). Far less popular is the right-hand side plot of Fig. \ref{rtxt} which shows the de-trended plot of returns, $x_t = r_t - \mu_1 t$, because fluctuations of $x_t$ are a result of market volatility. Despite an obviously very complicated nature of volatility, one of its features is remarkably simple: mean realized variance $\left<\mathrm{d}x_t^2\right>$ (or, equivalently, $m_2(\tau)$ -- variance of distribution of stock returns) depends linearly on the number of days $\tau$ of returns accumulation. This can be clearly seen in Figs. \ref{RV} from \cite{liu2019distributions} and \ref{RV2} from \cite{farahani2025asymmetry}, as well as Fig. \ref{m2tau} below. Notwithstanding a very different index composition, DJIA exhibits a very similar behavior \cite{liu2019distributions}. 
%\begin{figure}[htbp]
   % \centering
   % \includegraphics[width=0.77\linewidth]{logsts0_tau_4plots_n1_det}
    % \caption{S\&P500; $x_t = r_t - \mu_1 t$ where index in $\mu_1$ reflects daily increments of $t$ ($\tau =1$ in text).}
    %\label{xt}
%\end{figure}
\begin{figure}[!htbp]
\centering
\includegraphics[width = 0.49 \textwidth]{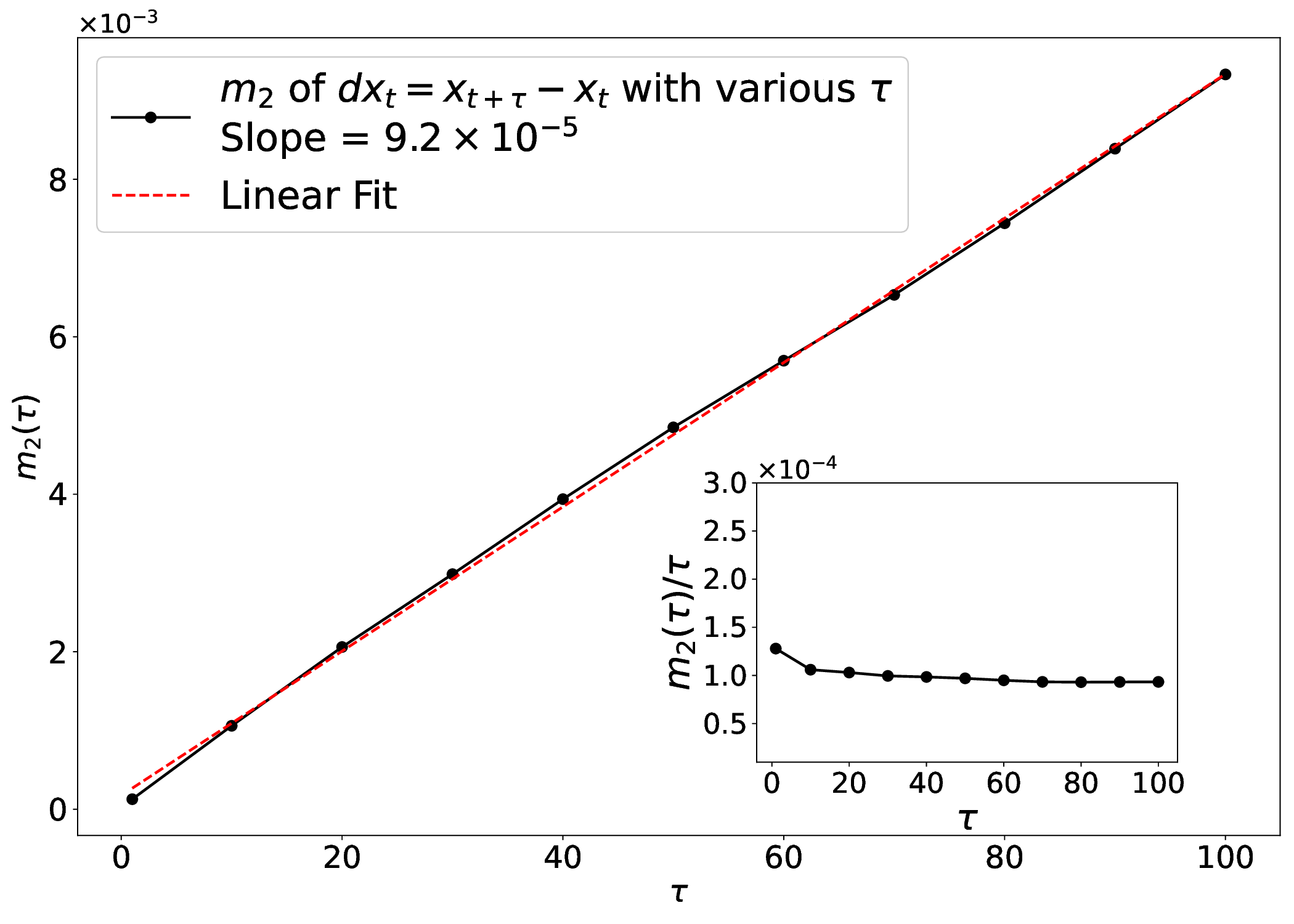}
\caption{The variance of the distribution of returns as a function of the number of days of accumulation, $m_2(\tau)$, with its linear fit. Inset shows scaled variance $m_2(\tau)/\tau$. From \cite {farahani2025asymmetry}.}
\label{RV2}
\end{figure}

As already mentioned, market volatility is an extremely complex phenomenon, which depends on multitude of disparate factors at any given time. Nonetheless, it begat a variety of models that try to describe it in a concise form. One of the simplest ones is based on a pair of stochastic differential equations (SDE) 
\begin{equation}
\mathrm{d}x_t =  \log\left(\frac{S_{t+\mathrm{d}t}}{S_t}\right) - \mu_1 \mathrm{d}t = \sigma_t \mathrm{d}W^{(1)}
\label{dxt}
\end{equation}
\begin{equation}
\mathrm{d}v_t = -\gamma(v_t - \theta)\mathrm{d}t + g(v_t) \mathrm{d}W^{(2)}
\label{dvt}
\end{equation}
where $\sigma_t$ is the stochastic volatility, $v = \sigma_t^{2}$ is the stochastic variance, and $\mathrm{d}W = W\left(t+\mathrm{d}t\right) - W(t)$ is the normally distributed Wiener process, $\mathrm{d}W \sim \mathrm{N(}0,\, \mathrm{d}t \mathrm{)}$, $(\mathrm{d}W)^2 = \mathrm{d}t$. In what follows we will neglect correlations between $\mathrm{d}W^{(1)}$ and $\mathrm{d}W^{(2)}$ since here we study distributions of returns which, unlike, for instance, leverage \cite{perello2003stochastic,dashti2021distributions}, seem to be unaffected by such correlations \cite{dragulescu2002probability, liu2019distributions}. 

The mean-reverting process for stochastic variance (\ref{dvt}) implies that $\left<v_t\right>$ relaxes to $\theta$ over $\gamma^{-1}$ timescale. In turn, it follows from (\ref{dxt}) that, for returns accumulated over $\tau$ days, $\mathrm{d}t= \tau$, the mean realized variance of returns should depend linearly on $\tau$
\begin{equation}
\left<\mathrm{d}x_t^2\right>=\theta \tau
\label{dxt2}
\end{equation}
which is in agreement with aforementioned empirical evidence. Furthermore, (\ref{dxt2}) does not depend on any particular choice of $g(v_t)$ and, thus, on the resultant steady-state (stationary) distributions of $v_t$ and $\sigma_t$.

To further illustrate the formalism based on (\ref{dxt}) and (\ref{dvt}), we point to numerous models for $g(v_t)$, such as Cox-Ingersoll-Ross (Heston)\cite{cox1985theory,heston1993closed,dragulescu2002probability,liu2019distributions}, multiplicative \cite{praetz1972distribution,nelson1990arch,fuentes2009universal,liu2019distributions}, and the combination of the two (multiplicative-Heston model)\cite{dashti2021combined}. In multiplicative and multiplicative-Heston models, for instance, $g(v_t)$ is given by, respectively,
\begin{equation} 
g_M(v_t)=\kappa_M v_t \qquad\text{and} \qquad g_{MH}(v_t)=\sqrt{\kappa_M^2 v_t^2 + \kappa_H^2 v_t }
\label{gvt}
\end{equation}
which leads, upon replacing $\mathrm{d}x_t$ with $x$, to the following steady-state probability density functions (PDF) of returns for multiplicative \cite{praetz1972distribution,fuentes2009universal,liu2019distributions} and multiplicative-Heston \cite{dashti2021combined} models respectively: 
\begin{equation}
f_{M}(x)=
\frac{\Gamma(\frac{\alpha}{\theta} + \frac{3}{2})}{\sqrt{\pi}\Gamma(\frac{\alpha}{\theta}+1)}\frac{1}{\sqrt{2\alpha \tau}}\left( \frac{x^2}{2\alpha \tau} + 1\right)^{-(\frac{\alpha}{\theta} + \frac{3}{2})} \quad\text{with}\quad \alpha = \frac{2\gamma \theta}{\kappa_M^2}  \quad\text{(Student t-distribution)}
\label{fM}
\end{equation}
\begin{equation}
f_{MH}(x) = \frac{\Gamma \left(q+\frac{1}{2}\right) U\left(q+\frac{1}{2}, \frac{3}{2}-p, \frac{x^2}{2 \beta  \tau}\right)}{\sqrt{2\pi \beta \tau} B \left(p,q \right)} \quad\text{with} \qquad p=\frac{2 \gamma \theta}{\kappa_H^2}, \qquad q=1+\frac{2 \gamma}{\kappa_M^2}, \qquad \beta=\frac{\kappa_H^2}{\kappa_M^2}
\label{fMH}
\end{equation}
Here $\Gamma(x)$ is the gamma function, $B(p,q)$ is the beta function, and $U$ is the confluent hypergeometric function. A direct calculation for either $f_M(x)$ or $f_{MH}(x)$ yields $\left<x^2\right>=\theta \tau$, in agreement with (\ref{dxt2}).

Obviously both of these PDFs are even (symmetric with respect to $x \rightarrow -x$), which means that behavior of gains and losses is treated as exactly the same in this formalism \footnote{This symmetry is unaffected by shifting the center of the distribution via rewriting l.h.s of (\ref{dxt}) as $\mathrm{d}(x_t-\mu)$.\label{shift}}. Furthermore, rescaling the variable with $\sqrt{2 \alpha \tau}$ and $\sqrt{2 \beta \tau}$ respectively would collapse PDFs into a single PDF \cite{fuentes2009universal}. Another prominent feature of these distributions is that their tails exhibit scale-free power-law behavior, respectively $\propto \pm{x}^{-2\left(\frac{\alpha}{\theta}+\frac{3}{2}\right)}$ and $\propto \pm{x}^{-(q+1)}$. While power-law tails in returns are not universally agreed upon, there is a strong case for them at least for daily returns, while for accumulated returns power law may persist for a large portion of the tail (see e.g. \cite{farahani2025asymmetry} and below).

While the symmetry of distributions (\ref{fM}) and (\ref{fMH}) is explicit through $x^2$ variable, symmetry of the distribution of returns is a generic property of the formalism based on (\ref{dxt}) and (\ref{dvt}): Similarly to (\ref{dxt2}), it does not depend on the choice of $g(v_t)$ and, therefore, neither on the specifics of $\sigma_t$ distribution. This is because the distribution of $\mathrm{d}x_t$ is inherently an even function since it is a product distribution \cite{ma2014model} of positively defined $\sigma_t$ distribution and normal distribution, which is symmetric. The fact that the formalism based on (\ref{dxt}) and (\ref{dvt}) is limited to symmetrical distributions is the crux of the problem since the symmetry of distribution of returns is clearly broken for actual empirical data. Namely, it is observed that the distribution of S\&P500 returns has \cite {farahani2025asymmetry, shao2026broken}
\begin{itemize}
	\item positive mean
	\item negative skew
	\item greater number of points for gains than for losses
	\item slower power-law exponent for losses than for gains
\end {itemize}
The central issue therefore that we attempt to address is \emph{whether it is possible to reconcile the linear dependence of realized volatility with asymmetrical distribution of returns, both of which observed empirically}. 

The key element of our approach is based on breaking symmetry of Student t-distribution (\ref{fM}) via the modified Jones-Faddy Skew t-Distribution (mJF1) \cite{jones2001skew,jones2003skew, shao2026broken}. While at this point we are unaware of whether or not it is possible to derive it from the first principles, such as SDE formalism, mJF1 provides a solid explanation of the above features with only relatively minor modifications to the Student t-distribution. This manuscript can be considered as a sister manuscript to \cite{shao2026broken}, which extends the latter to multi-day accumulated returns. It is organized as follows. In Section \ref{mJF1} we provide analytical framework for mJF1 distribution. Section \ref{numerics} summarizes the results of our numerical simulations based on fitting S\&P500 returns with mJF1 distribution. We summarize and discuss our results in Section\ref{summary}.

\section{Modified Jones-Faddy Distribution mJF1\label{mJF1}}
PDF of the modified Jones-Faddy distributions (mJF1) introduced in \cite{shao2026broken} for fitting of distribution of returns is given by
\begin{equation}
f(x)=C\left(1-\frac{x-\mu}{\sqrt{(x-\mu)^2+(\alpha_g+\alpha_l) \tau}}\right)^{\frac{\alpha_g}{\theta}+\frac{3}{2}} \left(1+\frac{x-\mu}{\sqrt{(x-\mu)^2+(\alpha_g+\alpha_l) \tau}}\right)^{\frac{\alpha_l}{\theta}+\frac{3}{2}}
\label{fmJF1}
\end{equation}
where the normalization constant $C$ is given by
\begin{equation}
C=\frac{1}{2^{\frac{\alpha_l}{\theta}+1+\frac{\alpha_g}{\theta}} \text{B}(\frac{\alpha_l}{\theta}+1,\frac{\alpha_g}{\theta}+1)  } \frac{1}{\sqrt{(\alpha_g+\alpha_l)\tau}}
\label{CmJF1}
\end{equation}
In general, CDF for gains and losses can be defined, respectively, as
\begin{equation}
	F_g (x) = \int_{-\infty}^{x} f(y) \mathrm{d}y \quad\text{and}\quad 	F_l (x) = \int_{x}^{\infty} f(y) \mathrm{d}y 
\label{FgFl}
\end{equation}
where $f(x)$ is the PDF of returns and $F_g (\infty)=F_l (-\infty)=1$. Complementary CDF, CCDF, are then $1-F_{g,l}(x)$. From (\ref{fmJF1}) and (\ref{FgFl}) we obtain mJF1 CDF in terms of regularized incomplete beta function \cite{nist2025digital} $I(x;a,b)$ as
\begin{equation}
F_{g\_mJF1} (x) =I\left(1+\frac{x-\mu}{\sqrt{(x-\mu)^2+(\alpha_g+\alpha_l) \tau}}; \frac{\alpha_g}{\theta} + 1,\frac{\alpha_l}{\theta} + 1 \right)
\label{FmJF1g}
\end{equation}
and
\begin{equation}
F_{l\_mJF1} (x) =I\left(1-\frac{x-\mu}{\sqrt{(x-\mu)^2+(\alpha_g+\alpha_l) \tau}}; \frac{\alpha_g}{\theta} + 1,\frac{\alpha_l}{\theta} + 1\right)
\label{FmJF1l}
\end{equation}

Comparison of (\ref{fM}) with (\ref{fmJF1}) shows two elements that differentiate them. First, as in standard Student distribution with location parameter\cite{wolfram2025student}, a location parameter $\mu$ is introduced here. Obviously it does not affect (\ref{dxt}) since the variable can always be shifted by a constant (see footnote \ref{shift}). Second -- and crucial -- difference is the introduction of a skew (skew t-distribution \cite{jones2001skew,jones2003skew,shao2026broken}), via $\alpha_g$ and $\alpha_l$ here. In particular, power-law tails scale as $x^{-\left( 2 \alpha_g/\theta+3 \right)}$ at $+\infty$ and $x^{-\left( 2 \alpha_l/\theta+3 \right)}$ at $-\infty$. This breaks a construct based on (\ref{dxt}) and (\ref{dvt}) which treats volatility of gains and losses uniformly: substitution $\alpha_g=\alpha_l=\alpha$ in (\ref{fmJF1}) leads back to (\ref{fM}) with non-zero location parameter $\mu$. At this point we are unaware of an SDE-based or otherwise first-principles formulation which would result in the distribution (\ref{fmJF1}).

Mean, variance and mode of mJF1 are given, respectively, by \cite{shao2026broken}
\begin{equation}
m_1 =\mu +\sqrt{(\alpha_g+\alpha_l)\tau}\, 
	\mathrm{B}\!\left(\frac{\alpha_g}{\theta}+\frac{1}{2},\frac{1}{2}\right)\,
       \mathrm{B}\!\left(\frac{\alpha_l}{\theta}+\frac{1}{2},\frac{1}{2}\right)\,
        \frac{\dfrac{\alpha_l}{\theta}-\dfrac{\alpha_g}{\theta}}{2\pi}\\
\label{m1}
\end{equation}
\begin{equation}
m_2 = \theta\tau
\frac{(\alpha_g+\alpha_l)^2}{4\alpha_g\alpha_l}
+\frac{(\alpha_g+\alpha_l)(\alpha_g-\alpha_l)^2\tau}{4\theta^2}
\left[
\frac{\theta^2}{\alpha_g\alpha_l}
-\left(\frac{\pi}{
\mathrm{B}\!\left(\frac{\alpha_g}{\theta},\frac{1}{2}\right)
\mathrm{B}\!\left(\frac{\alpha_l}{\theta},\frac{1}{2}\right)}\right)^2
\right].
\label{m2}
\end{equation}
\begin{equation}
\overline{m}=\mu + \sqrt{(\alpha_g+\alpha_l)\tau} \frac{\dfrac{\alpha_l}{\theta}-\dfrac{\alpha_g}{\theta}}
{2\sqrt{\left(\dfrac{\alpha_g}{\theta}+\dfrac{3}{2}\right)
\left(\dfrac{\alpha_l}{\theta}+\dfrac{3}{2}\right)}},
\label{mbar}
\end{equation}
We use first and second Pearson coefficients of skewness
\begin{equation}
\zeta_1=\frac{\left(m_1-\overline{m}\right)}{m_2^{1/2}}, \qquad \zeta_2=\frac{\left(m_1-\widetilde{m}\right)}{m_2^{1/2}}
\label{zeta12}
\end{equation}
to characterize skewness of the distribution, where $\widetilde{m}$ is the median and $m_2^{1/2}$ is the standard deviation. Once parameters $\alpha_g, \alpha_l, \theta \hspace{0.25em} \text{and} \hspace{0.25em} \mu$ are obtained in Sec. \ref{numerics} through Bayesian fitting, $m_1, m_2 \hspace{0.25em} \text{and} \hspace{0.25em} \overline{m}$ are evaluated from (\ref{m1}), (\ref{m2}) and (\ref{mbar}). $\widetilde{m}$ is evaluated numerically from the fitted distribution (see also Appendix in \cite{shao2026broken}).

\section{Numerical Results\label{numerics}}

\subsection{Fitting Parameters \label{params}}

Table \ref{paramsfit} lists parameters of (\ref{fmJF1}) estimated from Bayesian fitting of S\&P500 returns accumulated over $\tau$ days. 
\begin{table}[h]
\centering
\renewcommand{\arraystretch}{1.25}
\caption{Parameter estimates from Bayesian fitting with (\ref{fmJF1}) of S\&P500 returns accumulated over $\tau$ days.}
\label{paramsfit}
\begin{tabular}{ccccc}
\toprule
$\tau$ & $\alpha_g$ & $\alpha_l$ & $\theta$ & $\mu$ \\
\midrule
1  & $7.92 \times 10^{-5}$  & $6.42 \times 10^{-5}$  & $1.42 \times 10^{-4}$  & $8.46 \times 10^{-4}$ \\
2  & $1.17 \times 10^{-4}$ & $8.74 \times 10^{-5}$  & $1.20 \times 10^{-4}$ & $2.28 \times 10^{-3}$ \\
3  & $1.32 \times 10^{-4}$  & $9.11 \times 10^{-5}$  & $1.14 \times 10^{-4}$  & $3.90 \times 10^{-3}$ \\
4  & $1.40 \times 10^{-4}$  & $9.40 \times 10^{-5}$ & $1.09 \times 10^{-4}$  & $5.03 \times 10^{-3}$ \\
5  & $1.38 \times 10^{-4}$  & $9.00 \times 10^{-5}$  & $1.06 \times 10^{-4}$  & $5.96 \times 10^{-3}$ \\
6  & $1.36 \times 10^{-4}$ & $8.74 \times 10^{-5}$ & $1.03 \times 10^{-4}$ & $6.77 \times 10^{-3}$ \\
7  & $1.35 \times 10^{-4}$ & $8.52 \times 10^{-5}$ & $1.01 \times 10^{-4}$ & $7.44 \times 10^{-3}$ \\
8  & $1.32 \times 10^{-4}$ & $8.11 \times 10^{-5}$ & $9.86 \times 10^{-5}$ & $8.30 \times 10^{-3}$ \\
9  & $1.29 \times 10^{-4}$ & $7.68 \times 10^{-5}$ & $9.66 \times 10^{-5}$ & $9.28 \times 10^{-3}$ \\
10 & $1.30 \times 10^{-4}$  & $7.52\times 10^{-5}$ & $9.55 \times 10^{-5}$ & $1.03 \times 10^{-2}$ \\
\bottomrule
\end{tabular}
\end{table}
The plots in \ref{gains} and \ref{losses} show the following for gains and losses respectively:
\begin{enumerate}
  \item On a log-log scale: CCDF of S\&P500 returns and CCDF, $1-F_{g_mJF1}$ and $1-F_{l_mJF1}$ per (\ref{FmJF1g}) and (\ref{FmJF1l}), with estimated parameters from Table \ref{paramsfit}.
  \item Tails of the above CCDF which now includes linear fit (LF) and its confidence interval (CI) \cite{janczura2012black}.
  \item p-values obtained by U-test for the tail points \cite{pisarenko2012robust}.
\end{enumerate}

U-test \cite{pisarenko2012robust} was initially developed to identify outliers such as Dragon Kings and negative Dragon kings \cite{sornette2012dragon} but, in general, can be viewed as a measure of goodness of fit \cite{liu2023dragon} in the tail region. Namely, p-values of points between $0.95$ and $0.05$ indicate that they belong to the fitted distribution with $95\%$ confidence level. Clearly, both LF and mJF1, which has power-law tails, do not seem to be consistent for larger $\tau$ with a tempered behavior of S\&P500 in the end tails of its distribution, which makes aforementioned linear dependence of the variance $m_2(\tau)$ of S\&P500 distribution on $\tau$ even more remarkable. This point is further discussed in Sec. \ref{stats} below. 

\begin{figure}[!htb]
    \centering
    \includegraphics[width=.7\linewidth]{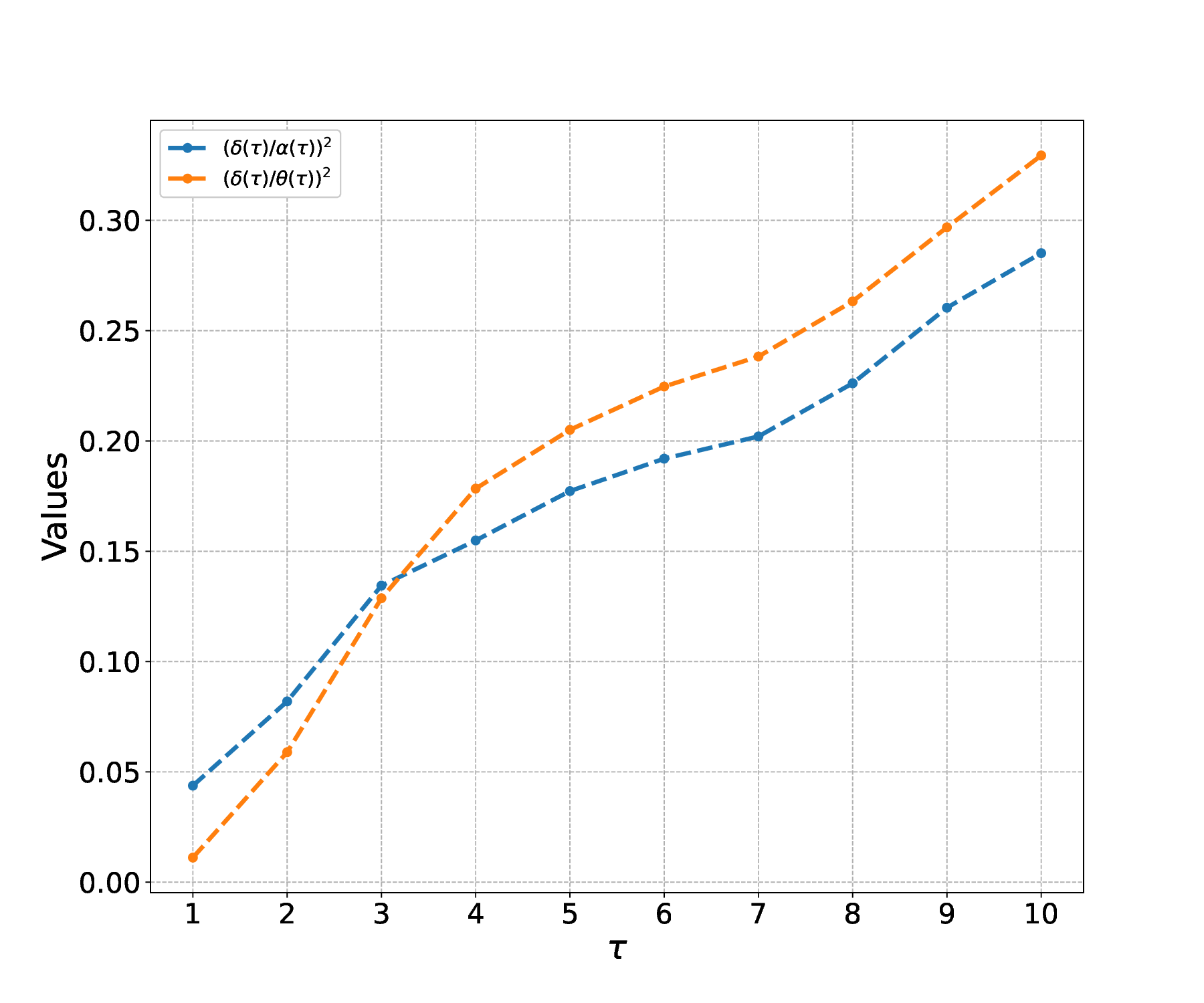}
    \caption{$(\delta/\alpha)^2$ and $(\delta/\theta)^2$ vs. $\tau$; $\delta=\alpha_g-\alpha_l$.}
    \label{delta}
\end{figure}

Fig. \ref{alphatheta} shows $\tau$ dependence of $\alpha_g$, $\alpha_l$ and $\theta$ from Table \ref{paramsfit}, as well as that of $\alpha=(\alpha_g+\alpha_l)/2$. Clearly, $\alpha$ and $\theta$ track each other closely after just a few days of accumulation. Fig. \ref{delta}, on the other hand, shows that ratios $\delta^2/\theta^2$ and $\delta^2/\alpha^2$, where $\delta=\alpha_g-\alpha_l$, track each other closely for all $\tau$. These ratios characterize the degree of divergence between tail exponents of gains and losses and will also be discussed in the context of $m_2(\tau)$ in Sec. \ref{stats} below. Fig. \ref{mu} shows $\tau$ dependence of the location parameter $\mu$ from Table \ref{paramsfit}. Its growth with $\tau$ seems to saturate to linear dependence and describes the shift of the bulk of the distribution to gains. This explains positive mean of the distribution while slower tails of losses explain negative skewness.

\begin{figure}[!htb]
    \centering
    \includegraphics[width=.7\linewidth]{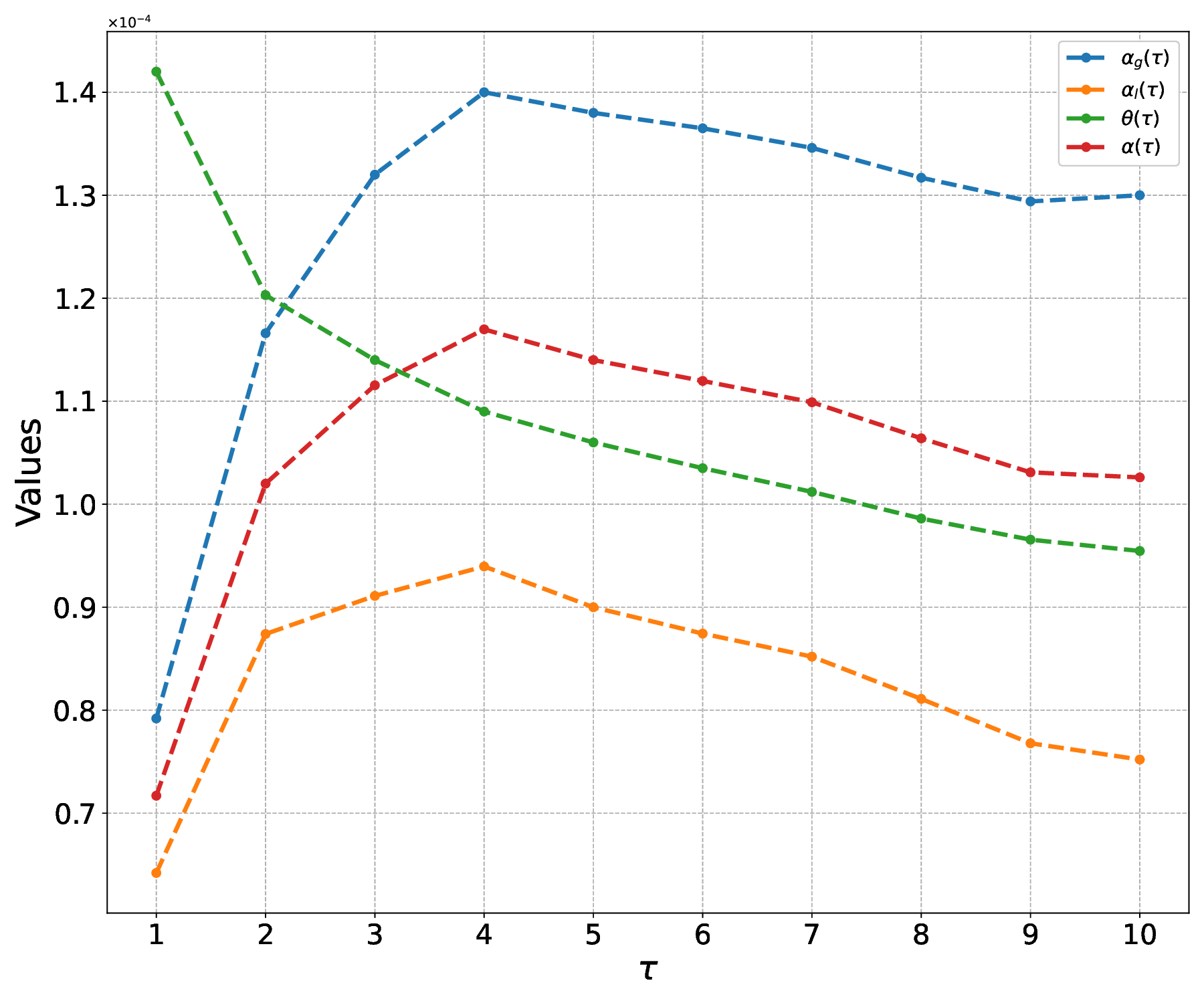}
    \caption{$\alpha_g$, $\alpha_l$ and $\theta$ from Table \ref{paramsfit} and $\alpha=(\alpha_g+\alpha_l)/2$ vs. $\tau$.}
    \label{alphatheta}
\end{figure}
\begin{figure}[!htb]
    \centering
    \includegraphics[width=.7\linewidth]{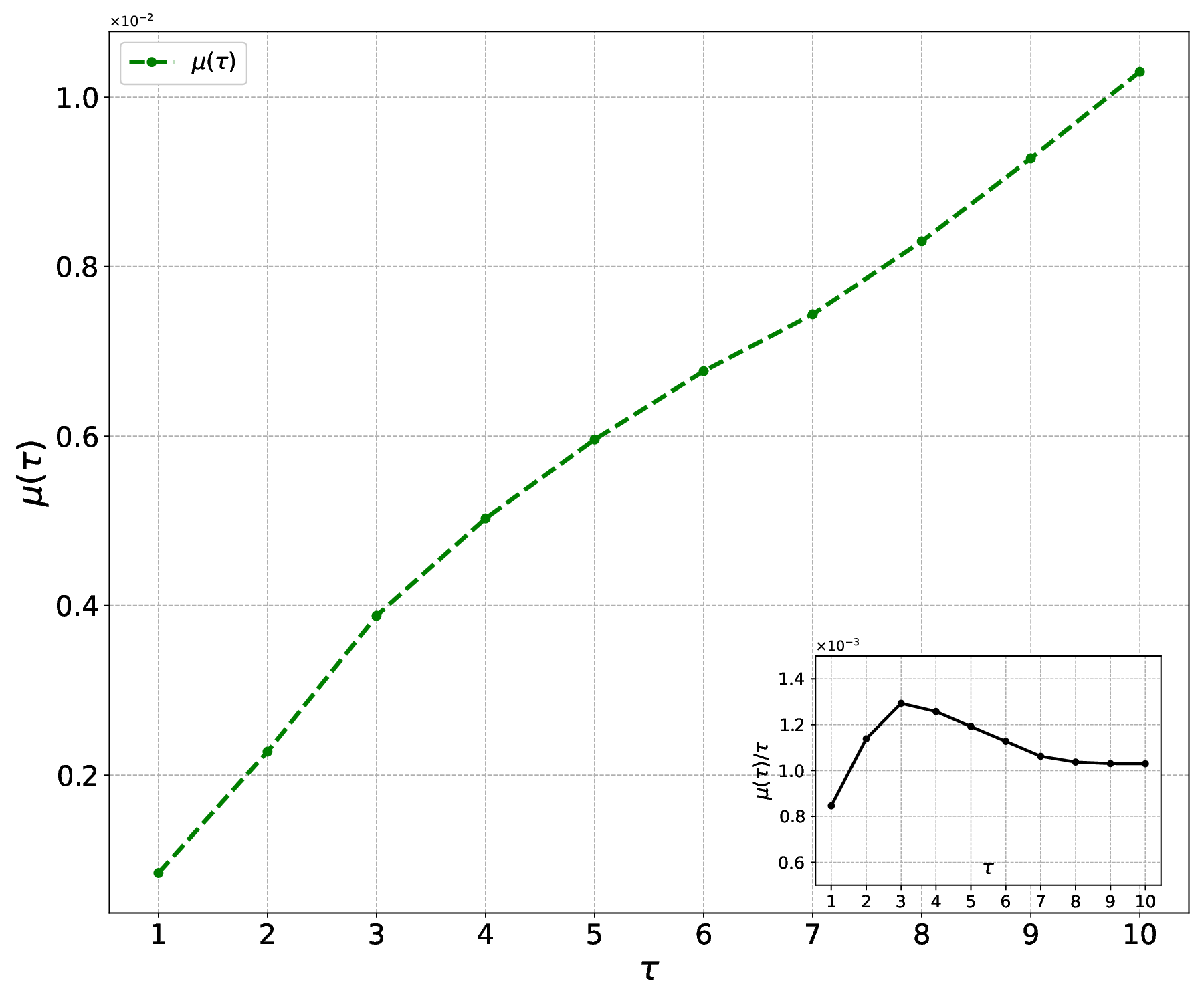}
    \caption{Location parameter $\mu$ from Table \ref{paramsfit} vs. $\tau$.}
    \label{mu}
\end{figure}

\subsection{Statistical Parameters \label{stats}}

\begin{figure}[!htb]
    \centering
    \includegraphics[width=.7\linewidth]{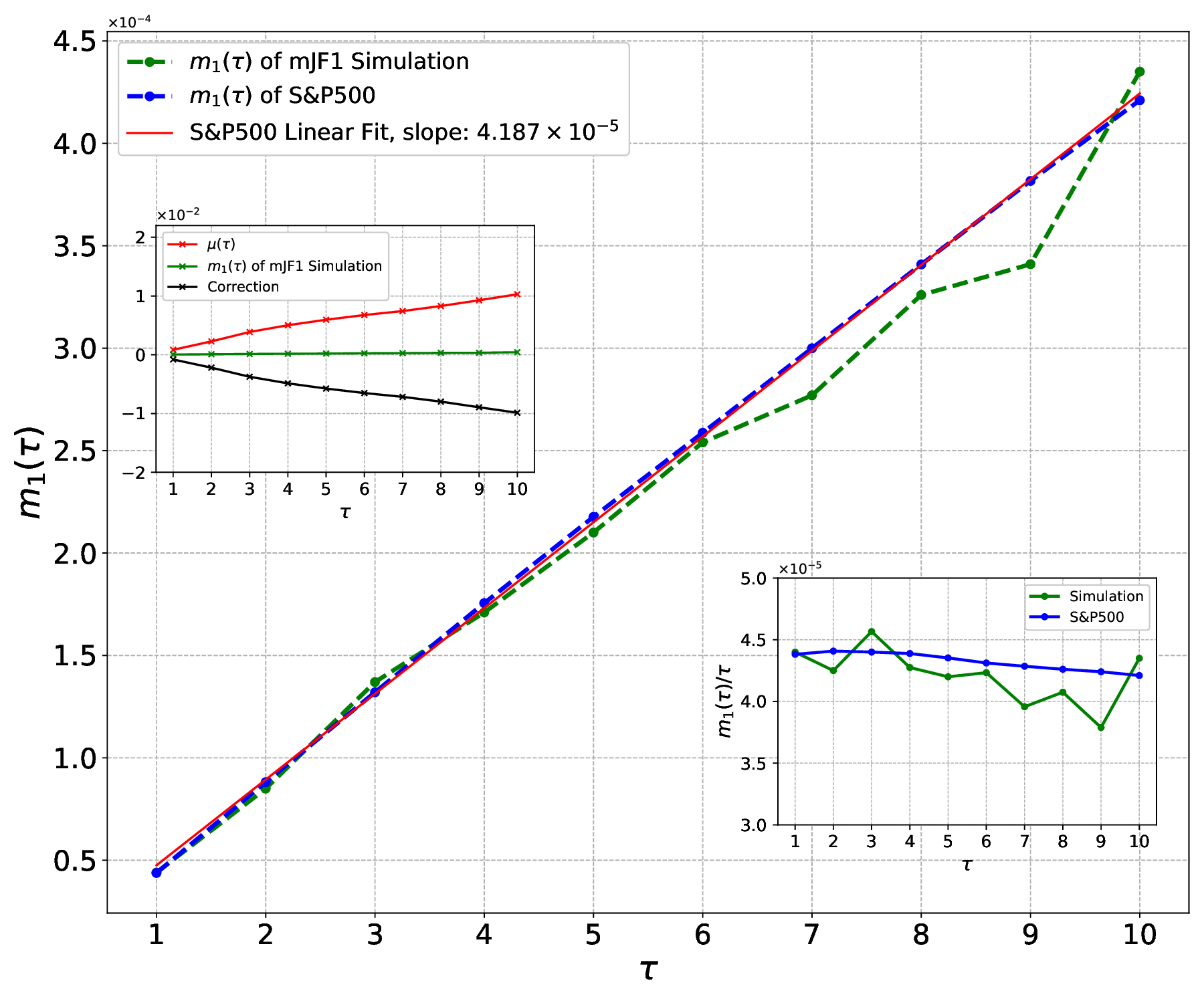}
    \caption{Mean $m_1(\tau)$ of S\&P500 and mJF1 fit and linear fit of S\&P500. Bottom insert: $m_1(\tau)/\tau$. Top insert: Positive and negative terms in (\ref{m1}) and resulting $m_1(\tau)$ -- notice two orders of magnitude difference between central plot and insert.}
    \label{m1tau}
\end{figure}

\begin{figure}[!htb]
    \centering
    \includegraphics[width=.7\linewidth]{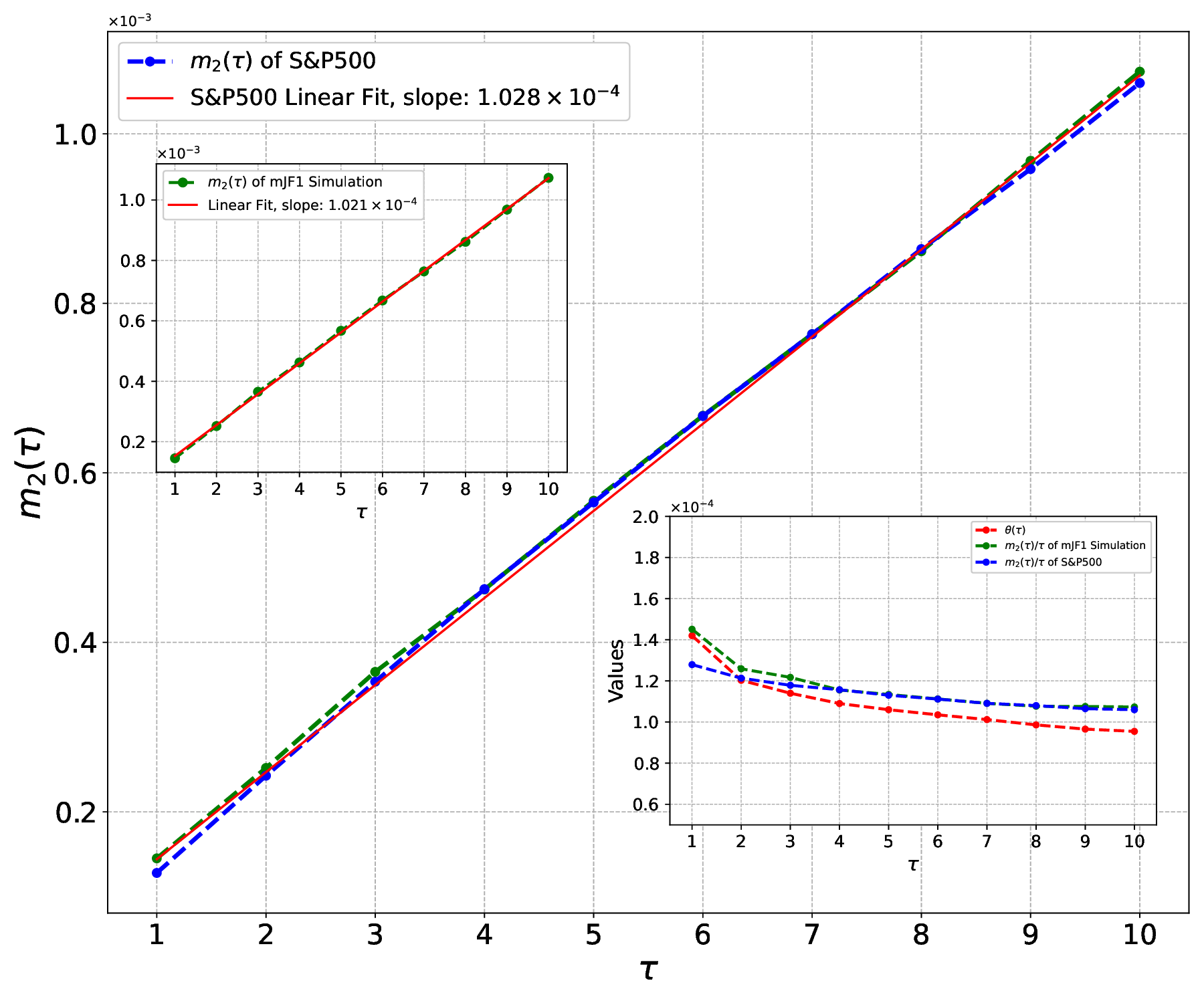}
    \caption{Variance $m_2(\tau)$ of S\&P500 and its linear fit. Top insert: $m_2(\tau)$ of mJF1 and its linear fit. Bottom insert: $m_2(\tau)/\tau$ of S\&P500 and mJF1 and $\theta(\tau)$.}
    \label{m2tau}
\end{figure}

Table \ref{m1m2} contains mean $m_1$ and variance $m_2$ of the S\&P500 distributions of returns, as well as those of fitted mJF1 distributions obtained using (\ref{m1}) and (\ref{m2}) with parameters of Table \ref{paramsfit}. Figs. \ref{m1tau} and \ref{m2tau} are graphical representation of $m_1(\tau)$ and $m_2(\tau)$ per Table \ref{m1m2}. Two very noteworthy features of Fig. \ref{m1tau} should be pointed out. First, $m_1(\tau)$ of mJF1 is a result of exquisite cancellation of two quantities in (\ref{m1}) which are orders of magnitude larger than $m_1(\tau)$: positive location parameter $\mu(\tau)$ discussed in Sec. \ref{params} and negative term due to slower decay of tails of losses, $\alpha_l<\alpha_g$. Second, mean value of S\&P500 distributions shows near perfect linear growth with $\tau$, which is well described by fitted mJF1.

\begin{figure}[!htb]
    \centering
    \includegraphics[width=.7\linewidth]{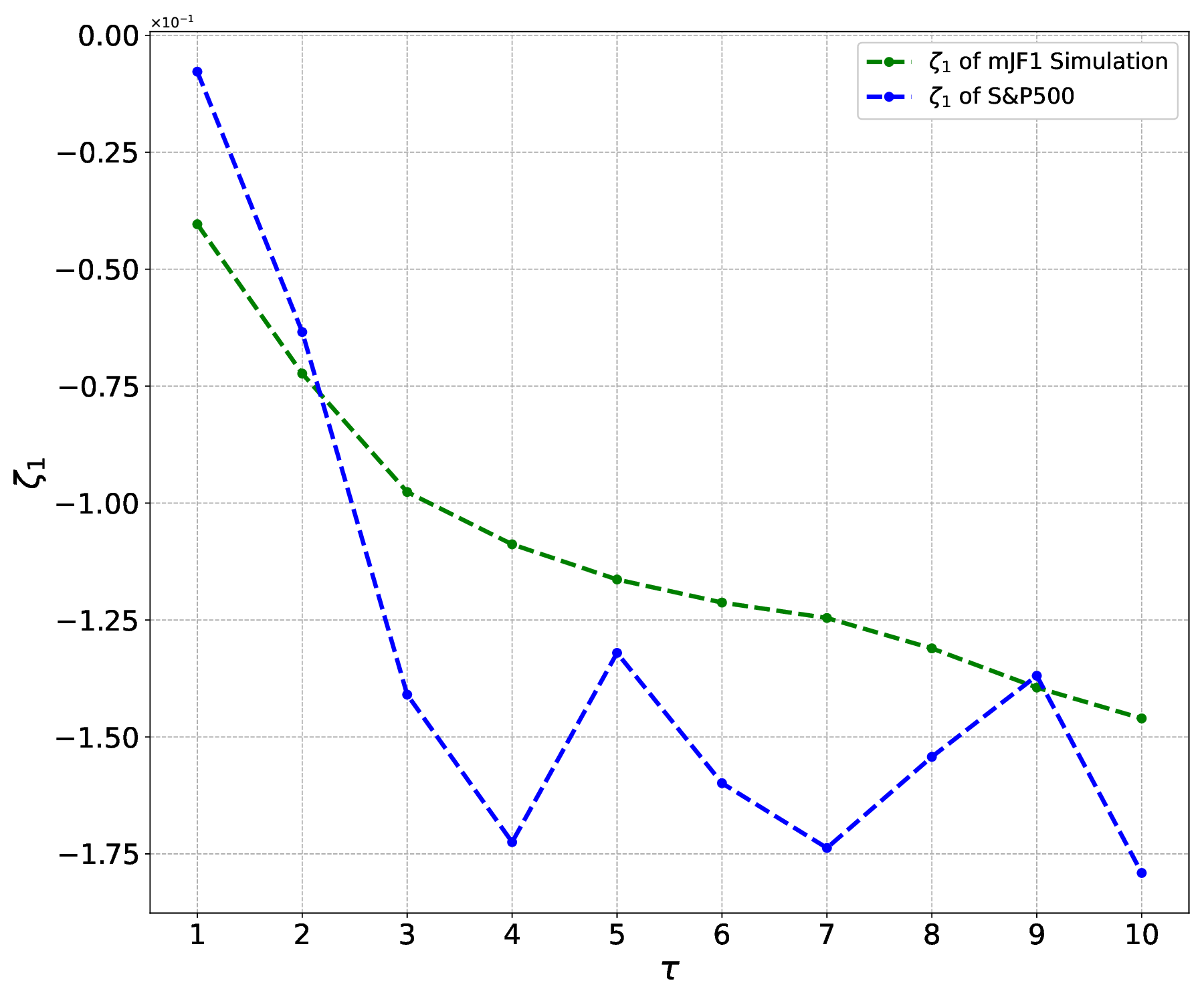}
    \caption{First Pearson coefficient of skewness $\zeta_1$ from Table \ref{skewness} vs. $\tau$.}
    \label{zeta1}
\end{figure}

\begin{figure}[!htb]
    \centering
    \includegraphics[width=.7\linewidth]{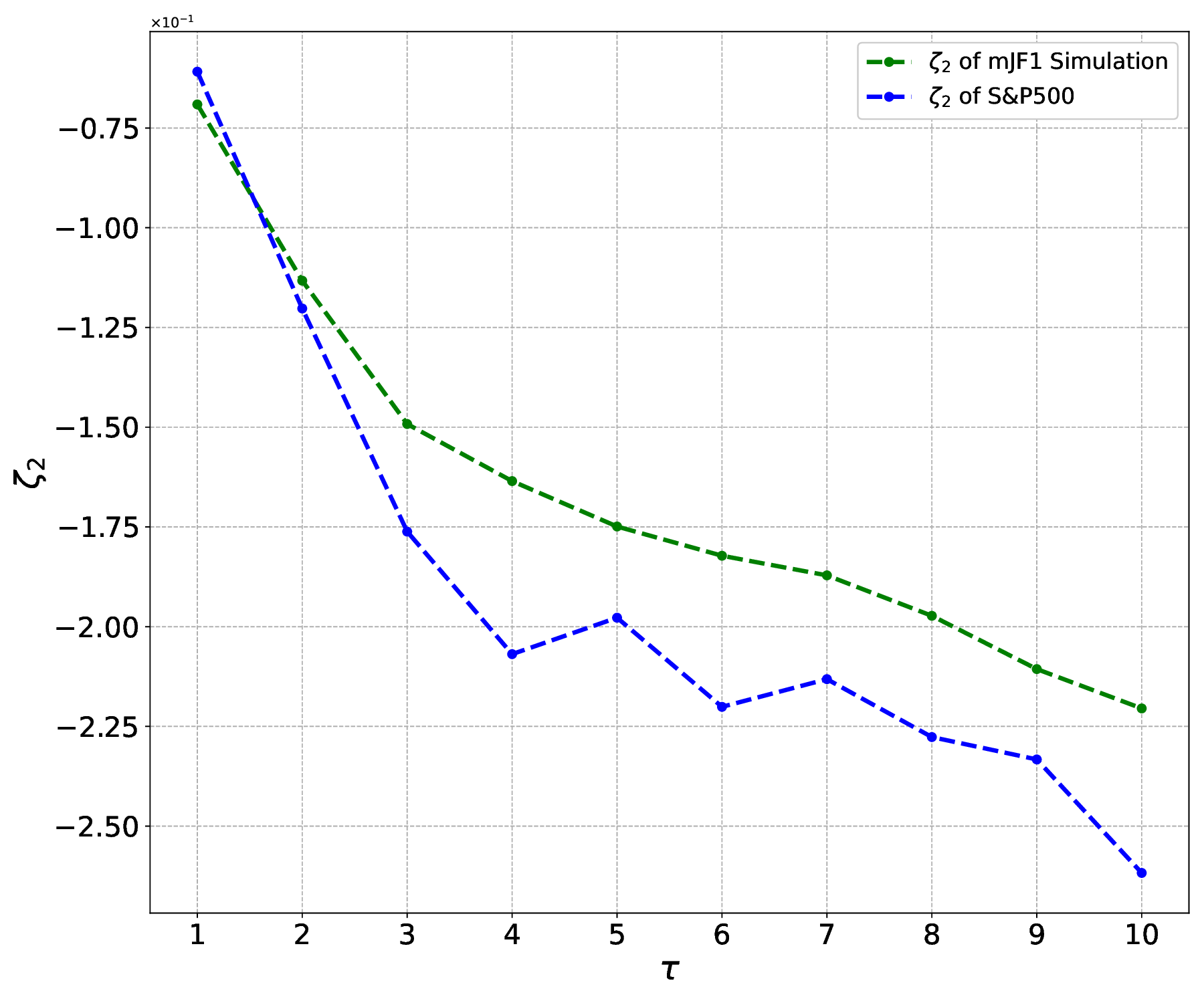}
    \caption{Second Pearson coefficient of skewness $\zeta_2$ from Table \ref{skewness} vs. $\tau$.}
    \label{zeta2}
\end{figure}

Linear dependence of $m_2(\tau)$ on $\tau$ is remarkably accurate both for S\&P500 and mJF1 fits, despite the former showing tempered behavior in the tail ends (see \ref{fit}) and the latter having power-law tails. This is expressly seen from the inserts in Fig. \ref{m2tau}: the top one shows that the slope of mJF1 $m_2(\tau)$ linear fit is nearly identical to that of S\&P500, and the bottom one shows that after four days of accumulation ratios $m_2(\tau)/\tau$ of the two are virtually indistinguishable and approach a saturation value. Additionally, the bottom insert contains a plot of $\theta(\tau)$, which is very close to $m_2(\tau)/\tau$ of S\&P500 and mJF1. The latter can be understood as follows. Assuming that $\delta = \alpha_g-\alpha_l \ll \alpha = (\alpha_g+\alpha_l)/2$, (\ref{m2}) simplifies to 
\begin{equation}
\frac{m_2(\tau)}{\theta\tau} \approx 1 + \frac{\delta^2}{4\alpha^2} \left[1 + \frac {2 \alpha}{\theta} - \frac{2\alpha^3}{\theta^3}  \frac{\Gamma^4 \left( \frac{1}{2}+\frac{\alpha}{\theta} \right)} {\Gamma^4 \left( \frac{\alpha}{\theta}\right)} \right]
\label{m2approx}
\end{equation}
While $\delta^2/\alpha^2$ (and $\delta^2/\theta^2$) are not necessarily very small, as seen from Fig. \ref{delta}, combination of other factors (coefficients, ratio $\alpha/\theta$, etc.) in (\ref{m2approx}) yields corrections to unity of 0.022 for $\tau=1$ and of 0.095 for $\tau=10$. The fact that $\theta$ tracks well  $m_2(\tau)/\tau$ gives support to mean-reverting volatility even in the mJF1 framework.
\begin{table}[h]
\centering
\renewcommand{\arraystretch}{1.25}
\caption{Mean $m_1$ and variance $m_2$ of mJF1 fits and of S\&P500.}
\label{m1m2}
\begin{tabular}{ccccc}
\toprule
$\tau$ & $m_1(\text{mJF1})$ & $m_1(\text{S\&P500})$ & $m_2(\text{mJF1})$ & $m_2(\text{S\&P500})$ \\
\midrule
1  & $4.39\times 10^{-5}$   & $4.38\times 10^{-5}$   & $1.45\times 10^{-4}$  & $1.28\times 10^{-4}$ \\
2  & $8.49\times 10^{-5}$   & $8.82\times 10^{-5}$   & $2.52\times 10^{-4}$  & $2.43\times 10^{-4}$ \\
3  & $1.37\times 10^{-4}$  & $1.32\times 10^{-4}$  & $3.65\times 10^{-4}$  & $3.54\times 10^{-4}$ \\
4  & $1.71\times 10^{-4}$  & $1.76\times 10^{-4}$  & $4.62\times 10^{-4}$  & $4.63\times 10^{-4}$ \\
5  & $2.10\times 10^{-4}$  & $2.18\times 10^{-4}$  & $5.67\times 10^{-4}$  & $5.65\times 10^{-4}$ \\
6  & $2.54\times 10^{-4}$  & $2.59\times 10^{-4}$  & $6.67\times 10^{-4}$  & $6.67\times 10^{-4}$ \\
7  & $2.77\times 10^{-4}$  & $3.00\times 10^{-4}$  & $7.64\times 10^{-4}$  & $7.64\times 10^{-4}$ \\
8  & $3.26\times 10^{-4}$  & $3.41\times 10^{-4}$  & $8.61\times 10^{-4}$  & $8.64\times 10^{-4}$ \\
9  & $3.41\times 10^{-4}$  & $3.82\times 10^{-4}$  & $9.68\times 10^{-4}$  & $9.58\times 10^{-4}$ \\
10 & $4.35\times 10^{-4}$  & $4.21\times 10^{-4}$  & $1.07\times 10^{-3}$ & $1.06\times 10^{-3}$ \\
\bottomrule
\end{tabular}
\end{table}
Table \ref{modemedian} gives mode $\overline{m}$ and median $\widetilde{m}$ of S\&P500 and mJF1. Finding mode of S\&P500 requires a smoothing procedure which may be behind a noticeable mismatch with the mode of mJF1. Mode and median are used to evaluate first and second Pearson coefficients of skewness via (\ref{zeta12}) whose values are given in Table \ref{skewness} and are illustrated in Figs. \ref{zeta1} and \ref{zeta2}.
\begin{table}[h]
\centering
\renewcommand{\arraystretch}{1.25}
\caption{Mode $\overline{m}$ and median $\widetilde{m}$ of mJF1 fits and of S\&P500.}
\label{modemedian}
\begin{tabular}{ccccc}
\toprule
$\tau$ & $\overline{m}(\text{mJF1})$ &$\overline{m}(\text{S\&P500})$ & $\widetilde{m}(\text{mJF1})$ & $\widetilde{m}(\text{S\&P500})$ \\
\midrule
1 & $5.30\times10^{-4}$ & $1.32\times10^{-4}$ & $3.21\times10^{-4}$ & $2.73\times10^{-4}$ \\
2 & $1.23\times10^{-3}$ & $1.08\times10^{-3}$ & $6.84\times10^{-4}$ & $7.13\times10^{-4}$ \\
3 & $2.00\times10^{-3}$ & $2.78\times10^{-3}$ & $1.09\times10^{-3}$ & $1.24\times10^{-3}$ \\
4 & $2.51\times10^{-3}$ & $3.89\times10^{-3}$ & $1.34\times10^{-3}$ & $1.66\times10^{-3}$ \\
5 & $2.98\times10^{-3}$ & $3.36\times10^{-3}$ & $1.60\times10^{-3}$ & $1.78\times10^{-3}$ \\
6 & $3.39\times10^{-3}$ & $4.39\times10^{-3}$ & $1.82\times10^{-3}$ & $2.15\times10^{-3}$ \\
7 & $3.72\times10^{-3}$ & $5.10\times10^{-3}$ & $2.00\times10^{-3}$ & $2.26\times10^{-3}$ \\
8 & $4.17\times10^{-3}$ & $4.87\times10^{-3}$ & $2.26\times10^{-3}$ & $2.57\times10^{-3}$ \\
9 & $4.68\times10^{-3}$ & $4.62\times10^{-3}$ & $2.53\times10^{-3}$ & $2.79\times10^{-3}$ \\
10 & $5.22\times10^{-3}$ & $6.25\times10^{-3}$ & $2.84\times10^{-3}$ & $3.26\times10^{-3}$ \\
\bottomrule
\end{tabular}
\end{table}
\begin{table}[h]
\centering
\renewcommand{\arraystretch}{1.25}
\caption{First and second Pearson coefficients of skewness, $\zeta_1$ and $\zeta_2$, of mJF1 fits and S\&P500.}
\label{skewness}
\begin{tabular}{ccccc}
\toprule
$\tau$ & $\zeta_1(\text{mJF1})$ & $\zeta_1(\text{S\&P500})$ & $\zeta_2(\text{mJF1})$ & $\zeta_2(\text{S\&P500})$ \\
\midrule
1 & $-4.04\times10^{-2}$ & $-7.76\times10^{-3}$ & $-6.91\times10^{-2}$ & $-6.09\times10^{-2}$ \\
2 & $-7.23\times10^{-2}$ & $-6.34\times10^{-2}$ & $-1.13\times10^{-1}$ & $-1.20\times10^{-1}$ \\
3 & $-9.76\times10^{-2}$ & $-1.41\times10^{-1}$ & $-1.49\times10^{-1}$ & $-1.76\times10^{-1}$ \\
4 & $-1.09\times10^{-1}$ & $-1.72\times10^{-1}$ & $-1.63\times10^{-1}$ & $-2.07\times10^{-1}$ \\
5 & $-1.16\times10^{-1}$ & $-1.32\times10^{-1}$ & $-1.75\times10^{-1}$ & $-1.98\times10^{-1}$ \\
6 & $-1.21\times10^{-1}$ & $-1.60\times10^{-1}$ & $-1.82\times10^{-1}$ & $-2.20\times10^{-1}$ \\
7 & $-1.24\times10^{-1}$ & $-1.74\times10^{-1}$ & $-1.87\times10^{-1}$ & $-2.13\times10^{-1}$ \\
8 & $-1.31\times10^{-1}$ & $-1.54\times10^{-1}$ & $-1.97\times10^{-1}$ & $-2.28\times10^{-1}$ \\
9 & $-1.39\times10^{-1}$ & $-1.37\times10^{-1}$ & $-2.10\times10^{-1}$ & $-2.33\times10^{-1}$ \\
10 & $-1.46\times10^{-1}$ & $-1.79\times10^{-1}$ & $-2.20\times10^{-1}$ & $-2.62\times10^{-1}$ \\
\bottomrule
\end{tabular}
\end{table}

\subsection{Scaling\label{scaling}}
\begin{figure}[!htb]
    \centering
    \includegraphics[width=.7\linewidth]{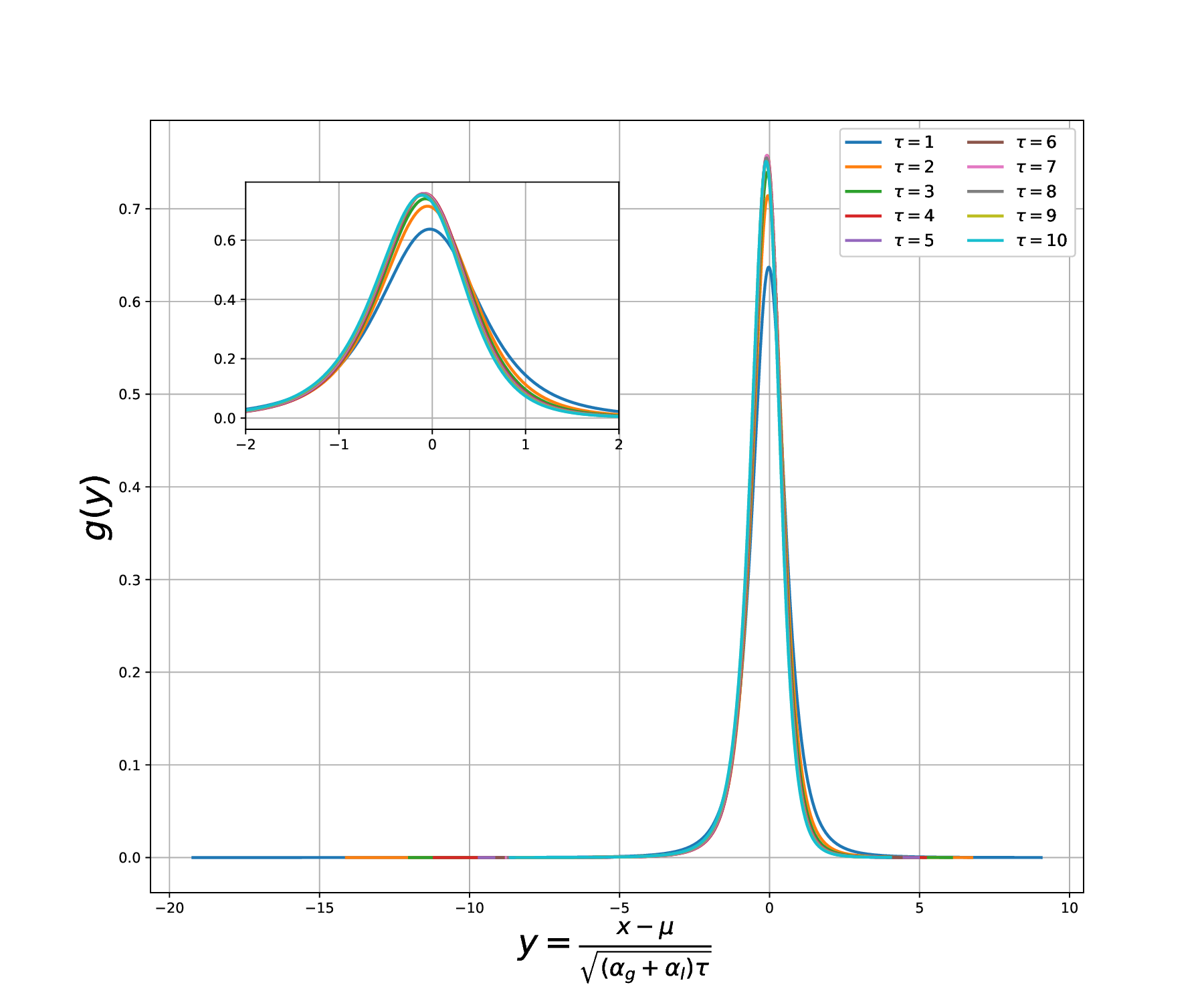}
    \caption{Rescaled PDF of mJF1.}
    \label{g}
\end{figure}

\begin{figure}[!htb]
    \centering
    \includegraphics[width=.7\linewidth]{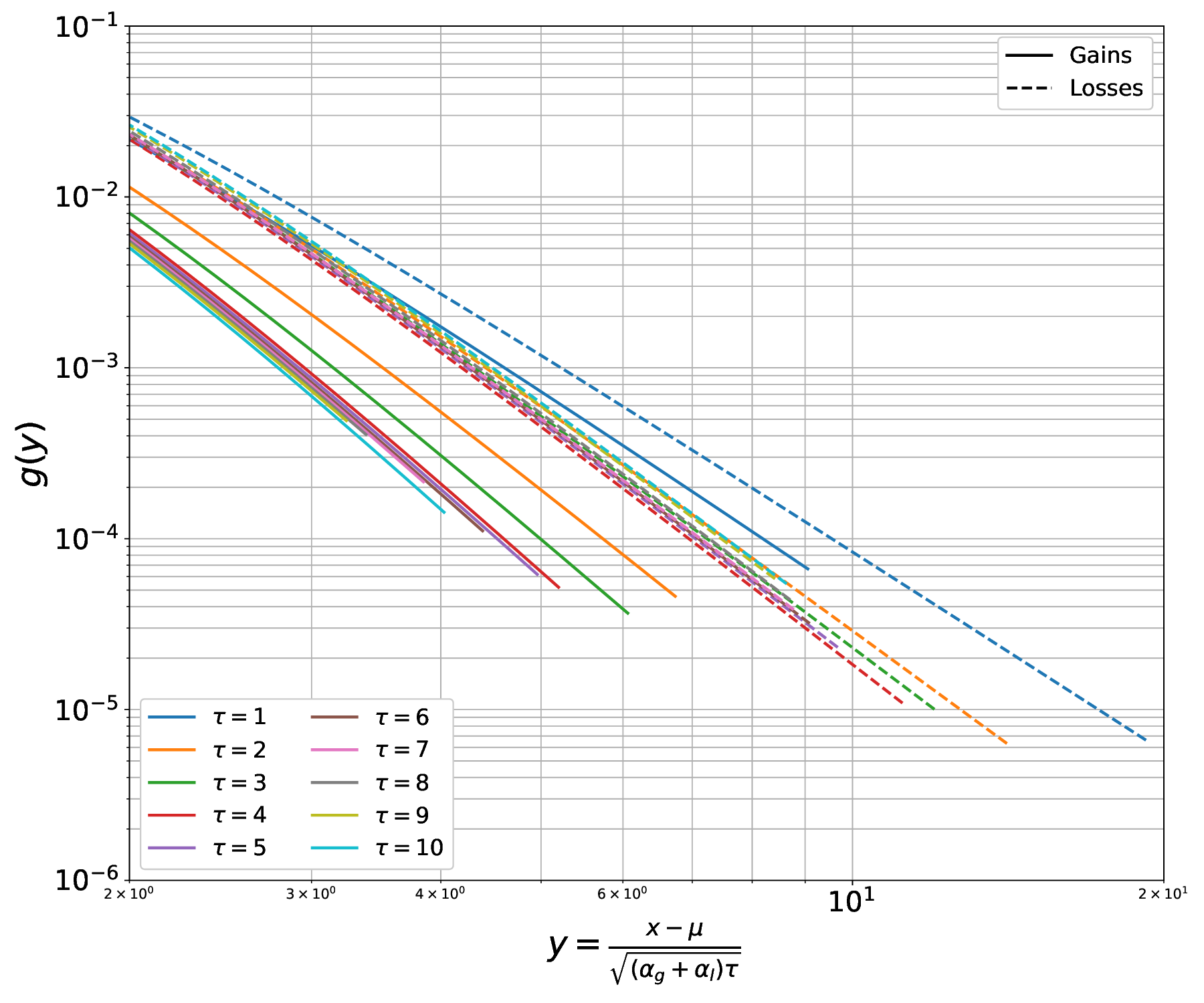}
    \caption{Tails of rescaled PDF of mJF1.}
    \label{gtails}
\end{figure}

The idea of scaling goes back to symmetrical models of distribution of stock returns \cite{praetz1972distribution,fuentes2009universal}. In particular, it is clear that a change of variable to $y=x/\sqrt{2\alpha\tau}$ in (\ref{fM}) and to $y=x/\sqrt{2\beta\tau}$ in (\ref{fMH}) removes dependence on $\tau$, so in each case the set of distributions for different $\tau$ should theoretically collapse to a single distribution. Of course, in practicality parameters are fitted for each $\tau$ and, consequently, there are some variations with $\tau$ \cite{liu2019distributions,dashti2021combined}. Similarly, the same concept applies to mJF1 (\ref{fmJF1}) via a change of variable $y=(x-\mu)/\sqrt{(\alpha_g+\alpha_l)\tau}$. This is illustrated in Figs. \ref{g} and \ref{gtails} which show PDF of rescaled distributions $g(y)$ with fitted mJF1 parameters from Table \ref{paramsfit}.

\newpage

\section{Summary and Discussion\label{summary}}
Empirical evidence very strongly suggests that realized variance (squared realized volatility) of major stock indices S\&P500 and DJIA scales linearly with the number of days of accumulations of returns. The impetus for this work was to understand this phenomenon in terms of a stochastic model of returns. In particular, (\ref{dxt2}) implies that if stochastic variance is on average fixed such linearity can be satisfactorily explained. However, mean-reverting models of stochastic variance, which would assure such average behavior, yield  symmetrical distributions (with respect to location parameter, if non-zero) -- see for instance (\ref{fM}) for multiplicative and (\ref{fMH}) for multiplicative-Heston models respectively.

In reality, the symmetry of the distributions of returns is broken, with the bulk of returns moving to gains while the tails of losses decaying slower than gains, which results in positive mean and negative skewness of the distribution of returns. To account for this, we employed a skew version of Student t-distribution (\ref{fM}) -- a modified Jones-Faddy skew t-distribution (\ref{fmJF1}). Bayesian fitting of S\&P500 returns produced good agreement with statistical properties of empirical distributions. One point, however, deserves particular attention.

Expression (\ref{m2}) for the variance of the modified Jones-Faddy distribution, while proportional to $\tau$, contains a rather complicated prefactor. When $\alpha_g=\alpha_l$ it reduces to $m_2=\theta\tau$, that is to (\ref{dxt2}), which is also the explicitly calculated variance of the  multiplicative and multiplicative-Heston models. Remarkably, with parameters obtained from fitting, the full expression still gives linear dependence on $\tau$ which is extremely close to that of S\&P500 as seen in Fig. \ref{m2tau} and explained in text. We dubbed this effect "conservation law." Its other manifestation is the scaling phenomenon as seen in Figs. \ref{g} and \ref{gtails}.

Future work will address extension to much larger values of the days of accumulation $\tau$ and generalization of modified Jones-Faddy distribution that would explain empirically observed tempered tails. Application of our approach to other indices and to returns of specific companies is also of great interest. Finally, a more direct consideration of the distribution of realized variance and linear dependence of its variance is also in order.

%\section{Funding}
%This research received no external funding.
%\section{Author Contributions}
%Conceptualization, R.A.S.; formal analysis, R.A.S. and S.S.; investigation, A.G., S.S., and R.A.S.; data curation, A.G.;  writing---original draft preparation, R.A.S., A.G., and S.S.; writing---review and editing, R.A.S., A.G., and S.S.; visualization, A.G..

\section{Data Availability}
We obtained S\&P500 data at Yahoo! Finance. Our datasets are available upon request.

\section{Acknowledgments}
We used MathWorks Matlab for numerical work and Wolfram Mathematica for analytical calculations. Siqi Shao acknowledges support in part by The University of Cincinnati URC Graduate Support Program.
%\section{Conflicts of Interest}
%The authors declare no conflicts of interest.
 
\clearpage

\bibliography{mybib}

\clearpage

\appendix
\section{Bayesian fitting with mJF1 \label{fit}}

\subsection{Gains\label{gains}}
\begin{figure}[!htb]
    \centering
    \includegraphics[width=1\linewidth]{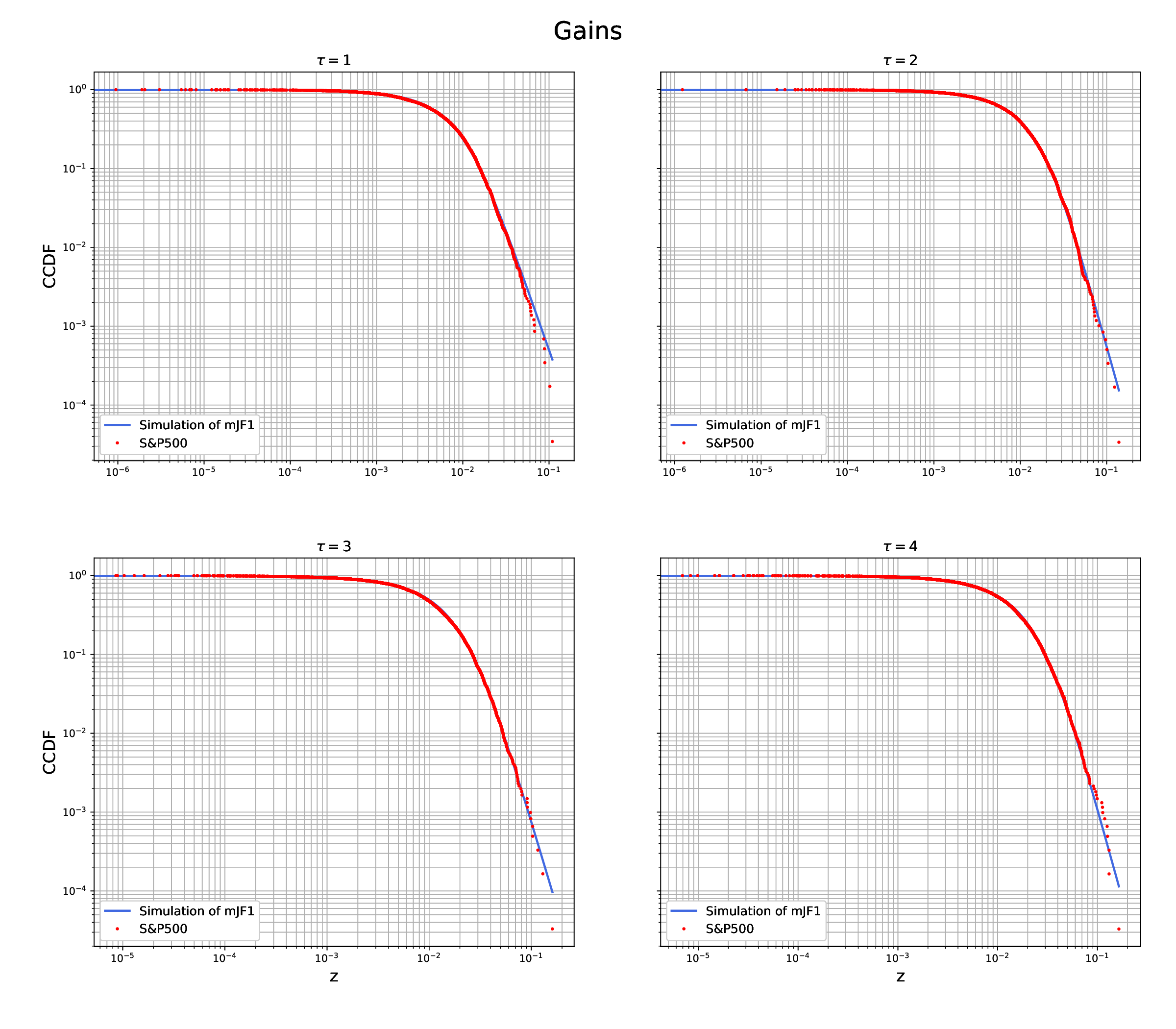}
    \caption{mJF1 CCDF}
    \label{g1-4}
\end{figure}

\begin{figure}[!htb]
    \centering
    \includegraphics[width=1\linewidth]{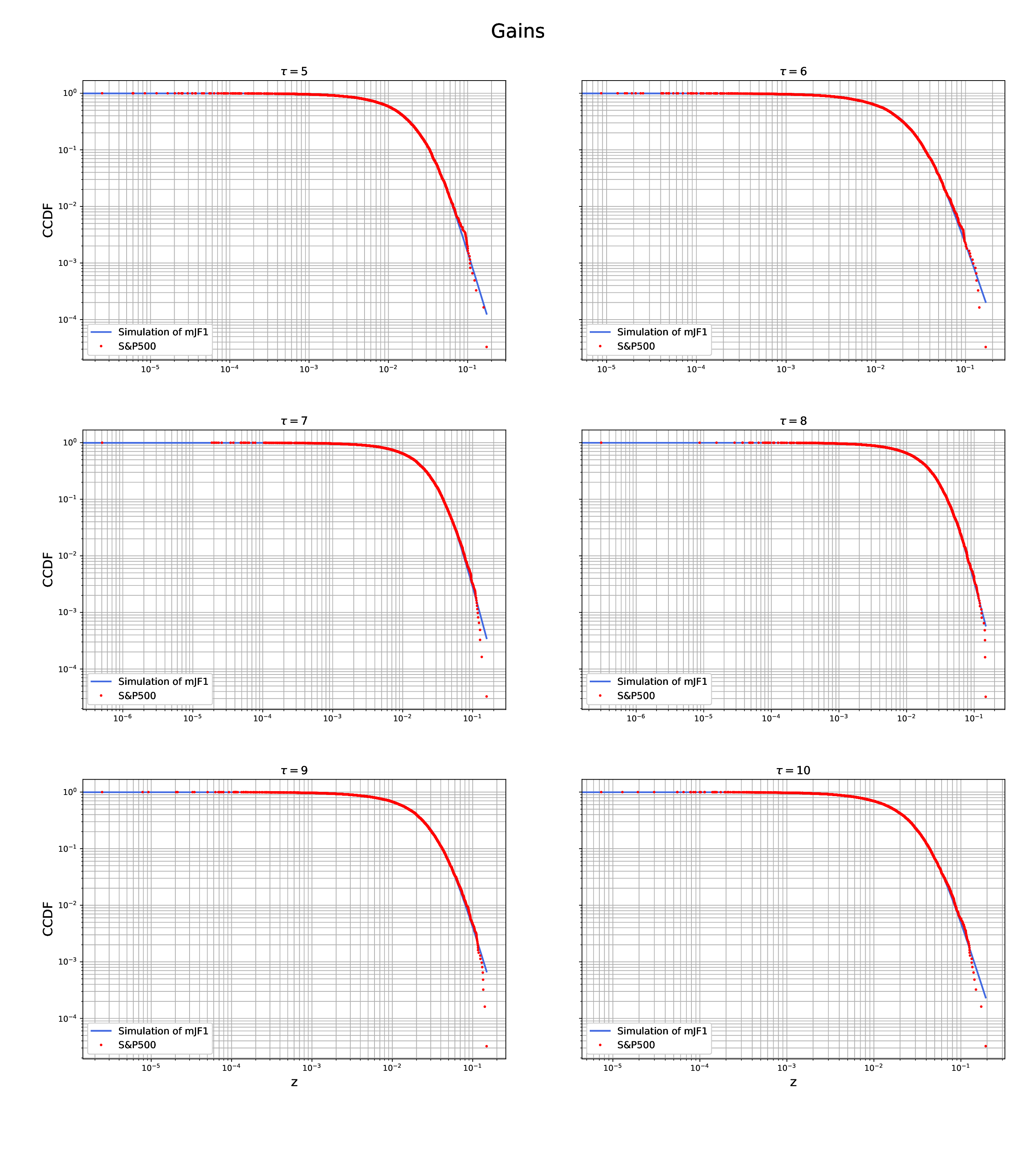}
    \caption{mJF1 CCDF}
    \label{g5-10}
\end{figure}

\begin{figure}[!htb]
    \centering
    \includegraphics[width=1\linewidth]{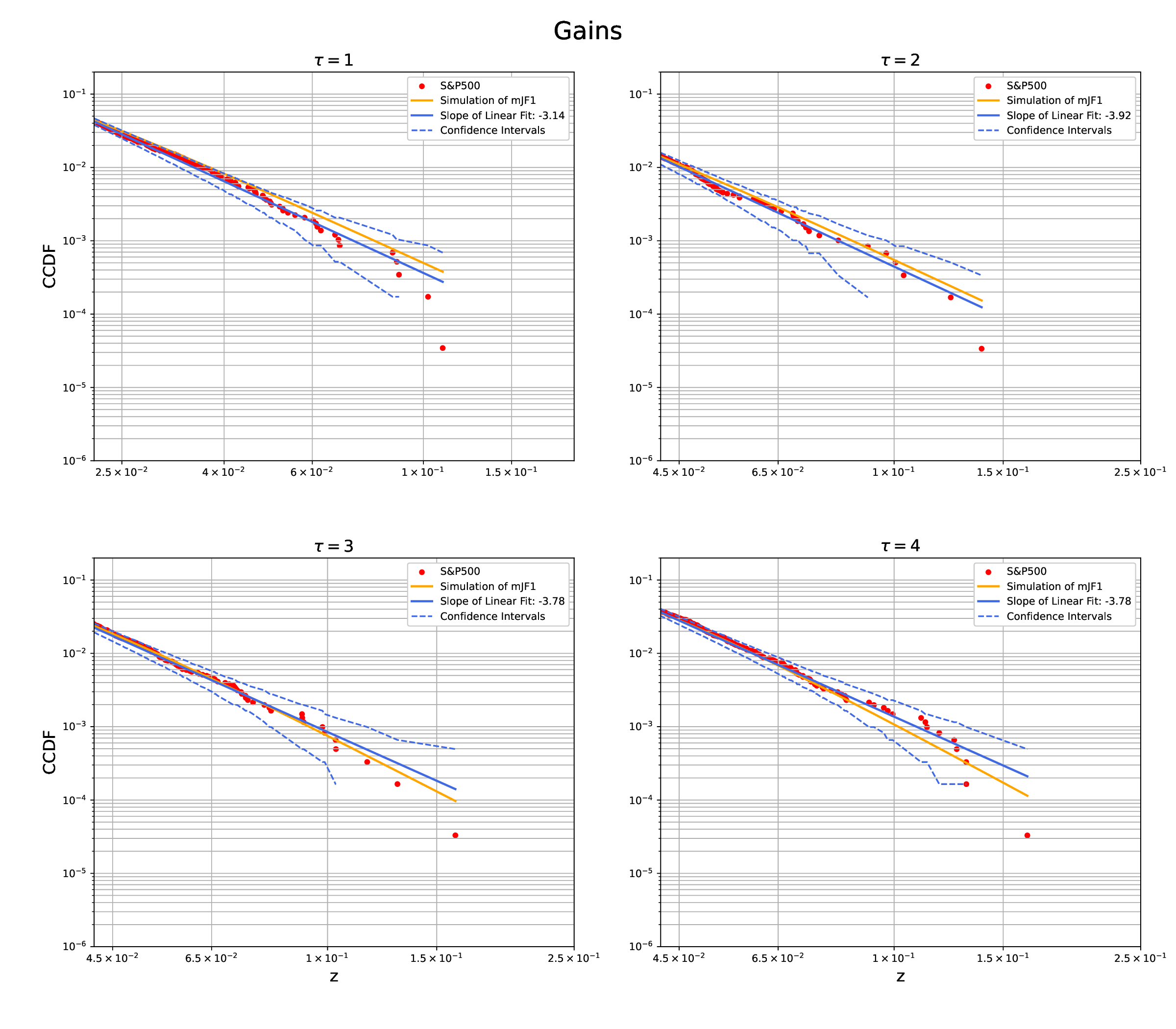}
    \caption{mJF1 CCDF Tails}
    \label{gtails1-4}
\end{figure}

\begin{figure}[!htb]
    \centering
    \includegraphics[width=1\linewidth]{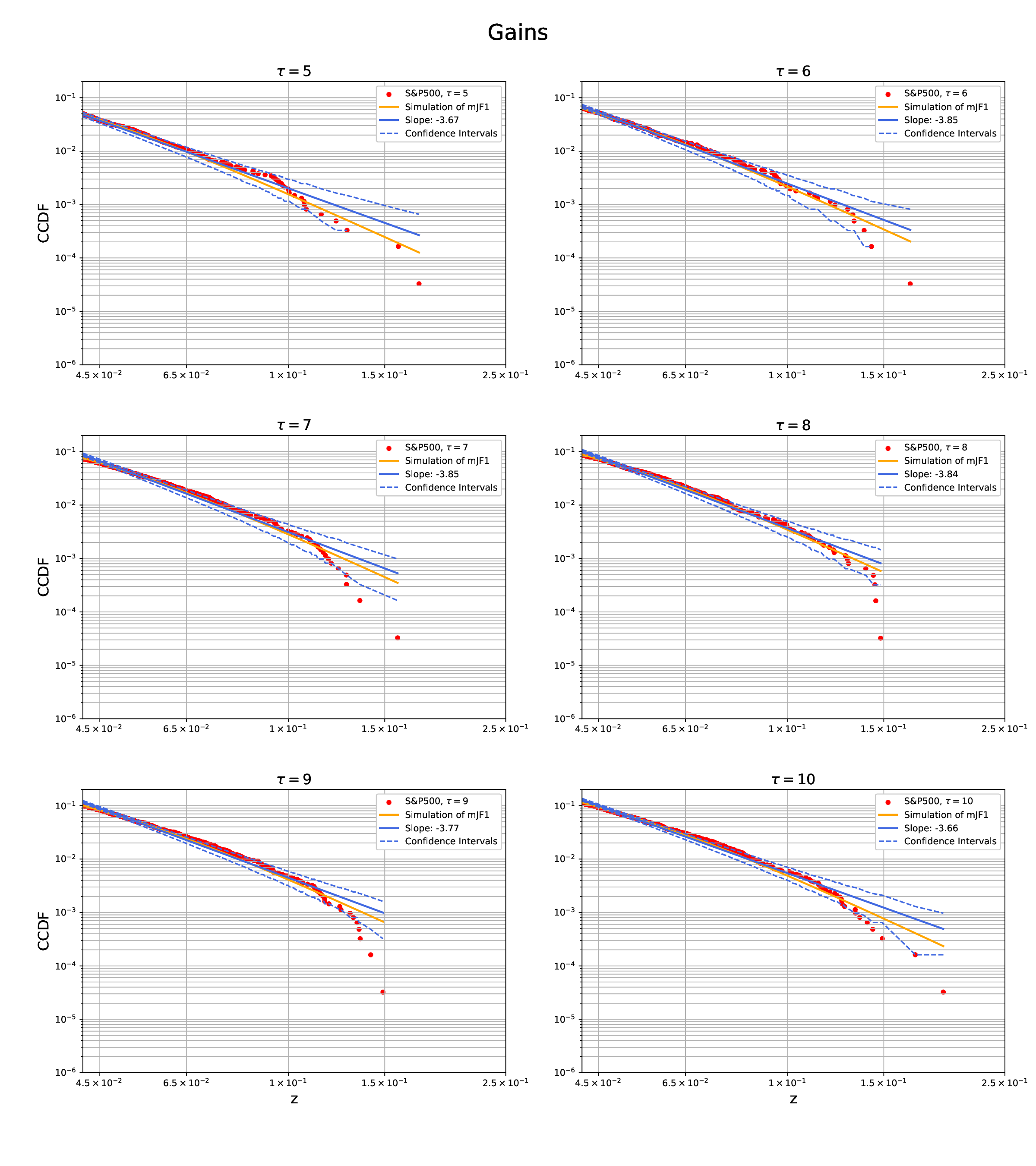}
    \caption{mJF1 CCDF Tails}
    \label{gtails5-10}
\end{figure}

\begin{figure}[!htb]
    \centering
    \includegraphics[width=1\linewidth]{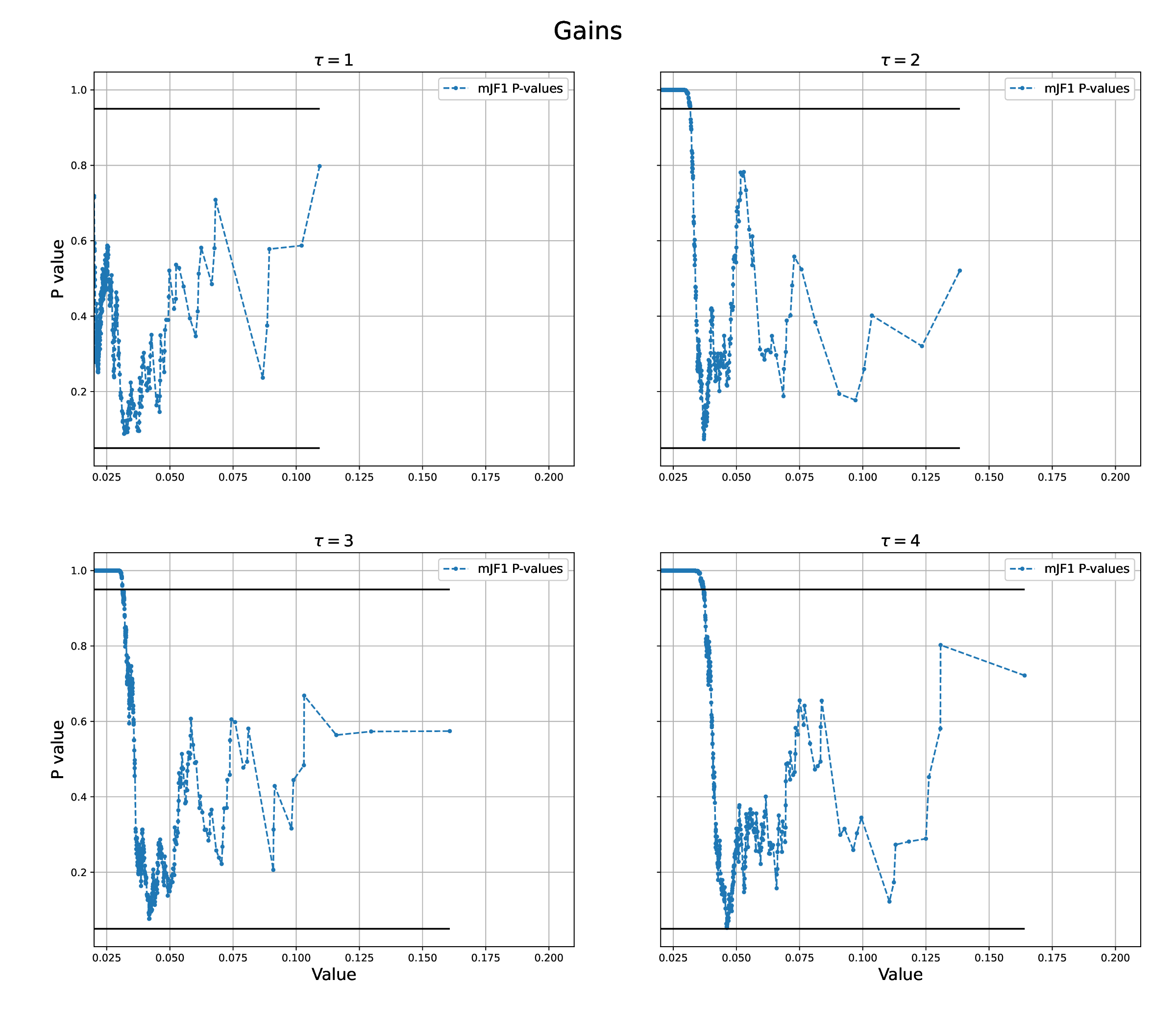}
    \caption{mJF1 p-values}
    \label{gp1-4}
\end{figure}

\begin{figure}[!htb]
    \centering
    \includegraphics[width=1\linewidth]{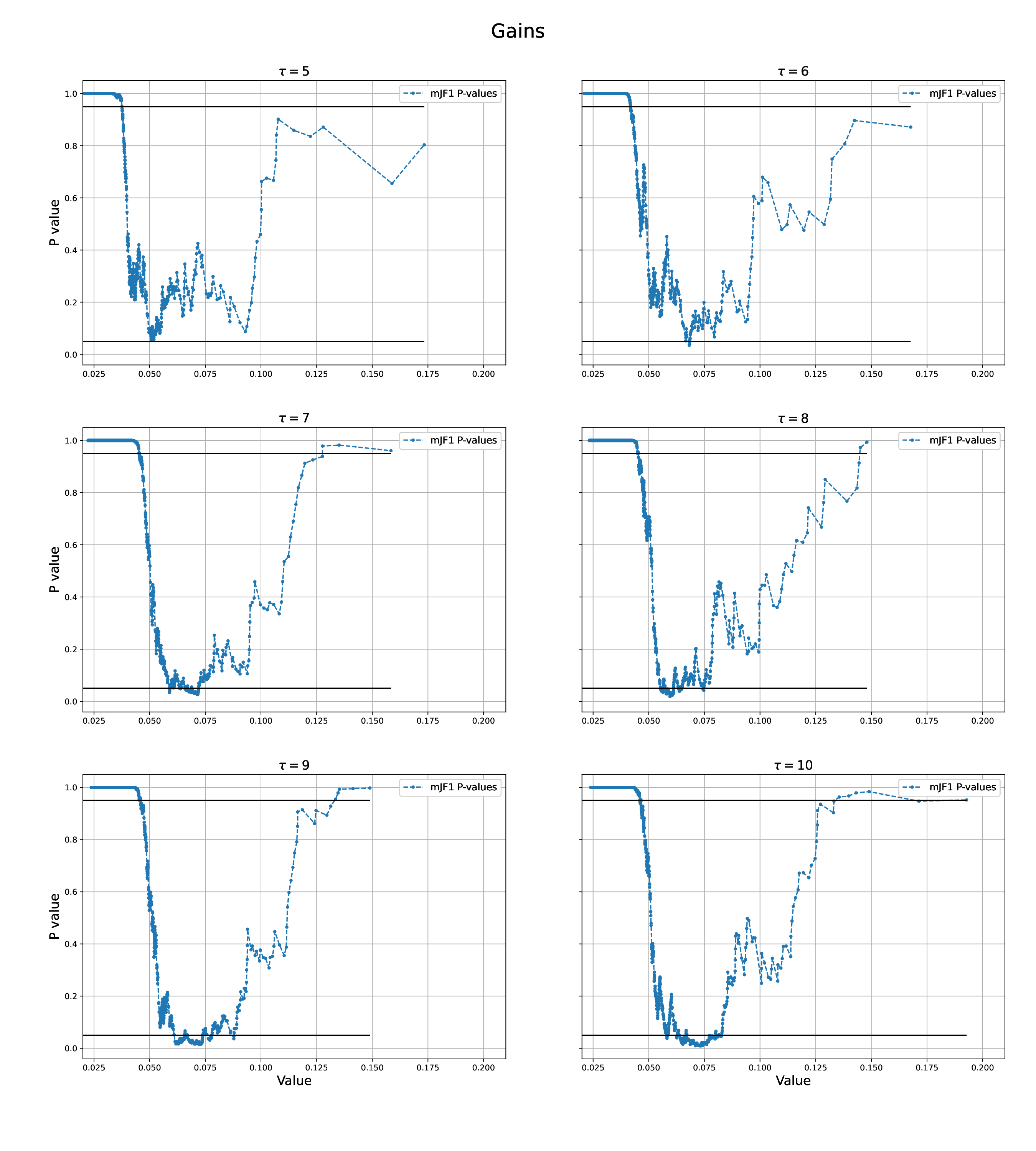}
    \caption{mJF1 p-values}
    \label{gp5-10}
\end{figure}

\clearpage

\subsection{Losses\label{losses}}

\begin{figure}[!htb]
    \centering
    \includegraphics[width=1\linewidth]{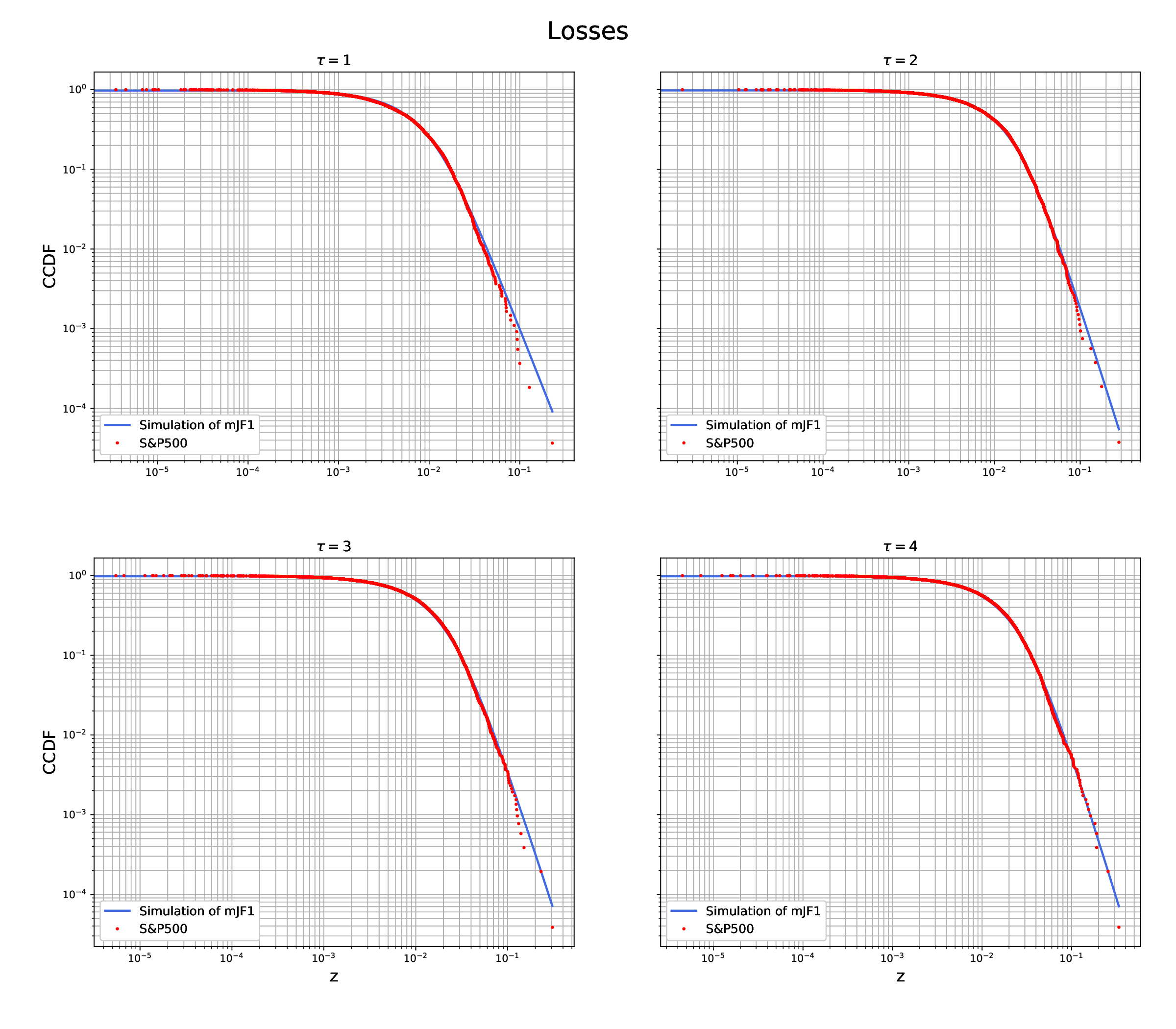}
    \caption{mJF1 CCDF}
    \label{l1-4}
\end{figure}

\begin{figure}[!htb]
    \centering
    \includegraphics[width=1\linewidth]{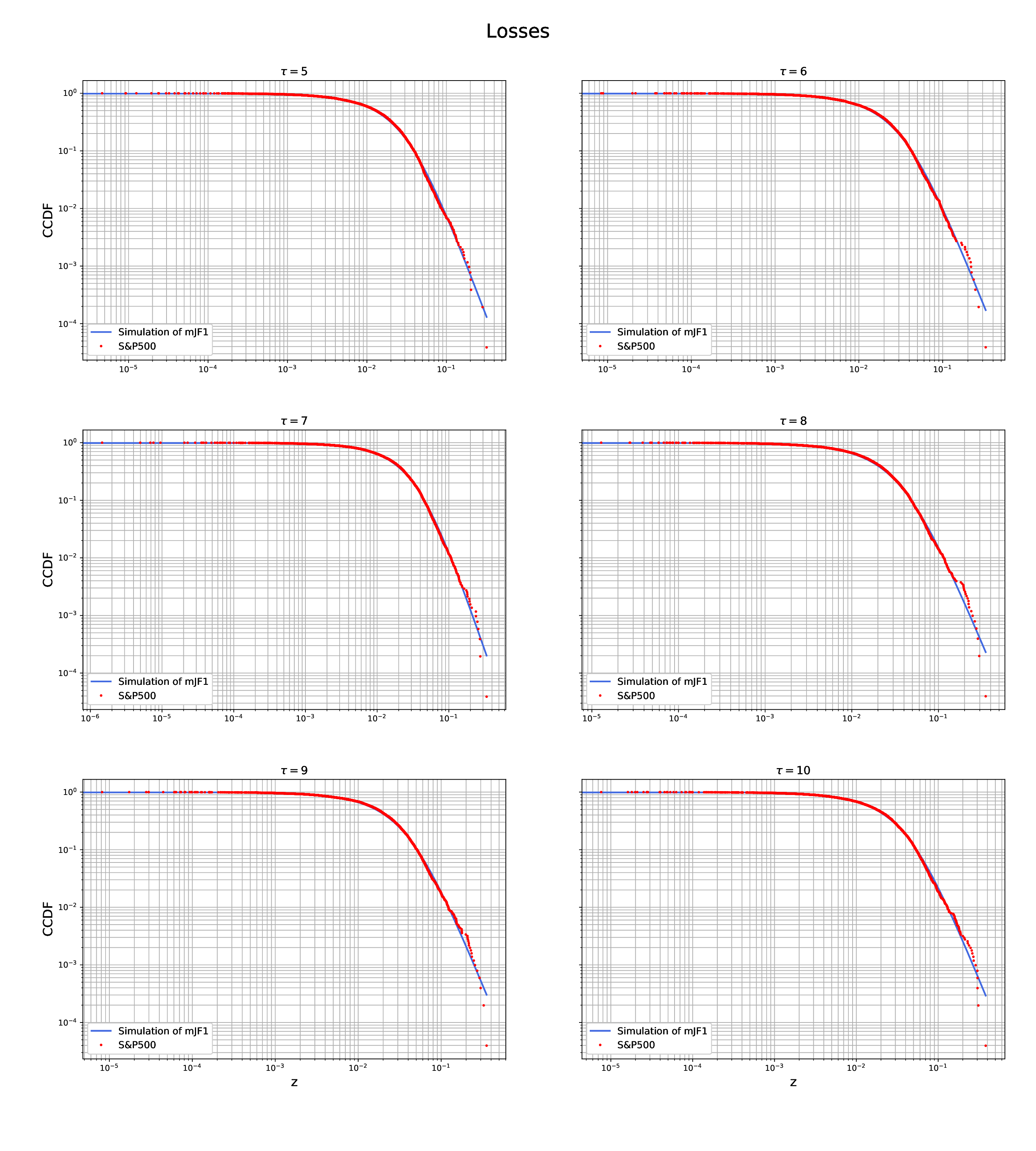}
    \caption{mJF1 CCDF}
    \label{l5-10}
\end{figure}

\begin{figure}[!htb]
    \centering
    \includegraphics[width=1\linewidth]{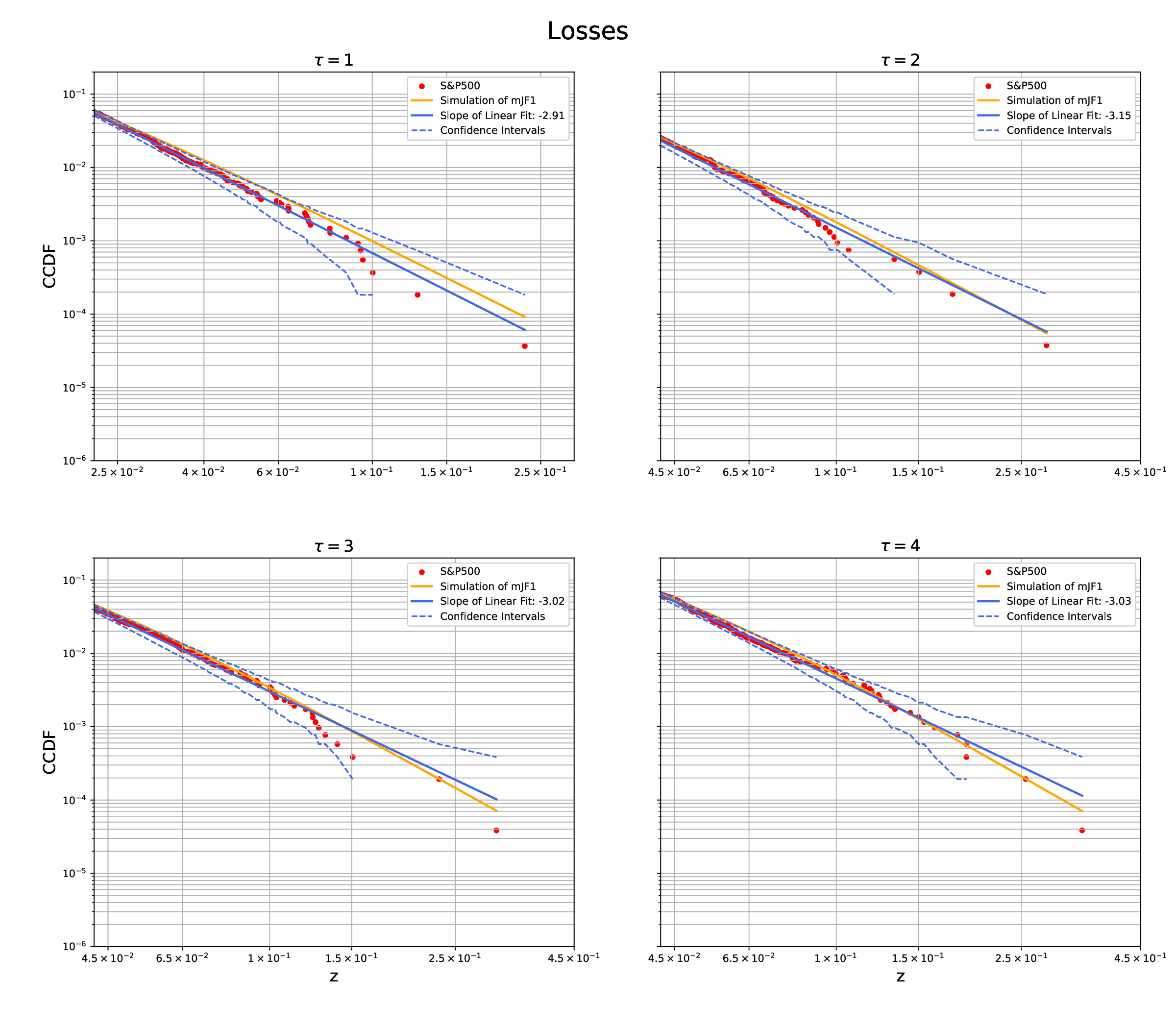}
    \caption{mJF1 CCDF Tails}
    \label{ltails1-4}
\end{figure}

\begin{figure}[!htb]
    \centering
    \includegraphics[width=1\linewidth]{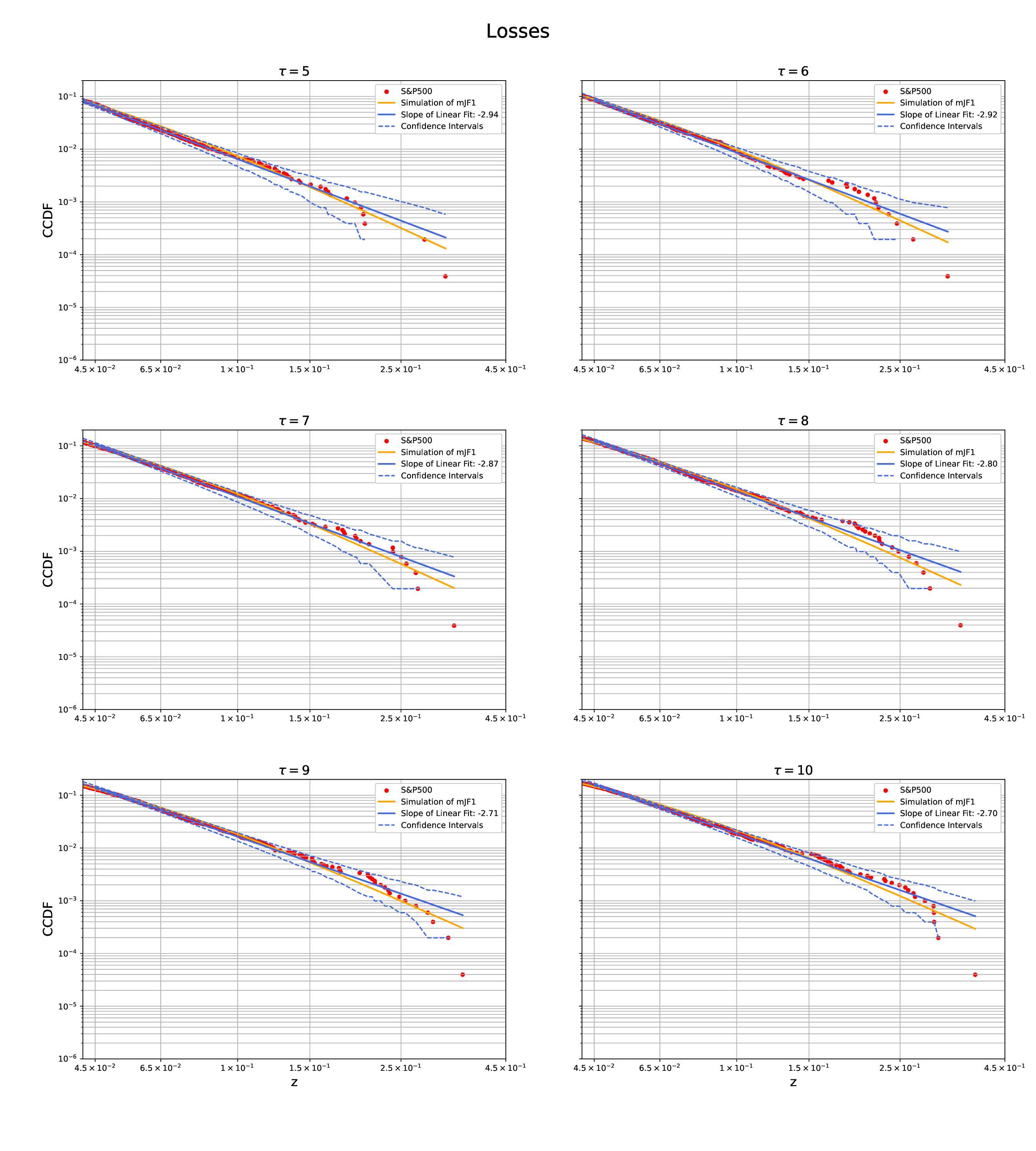}
    \caption{mJF1 CCDF Tails}
    \label{ltails5-10}
\end{figure}

\begin{figure}[!htb]
    \centering
    \includegraphics[width=1\linewidth]{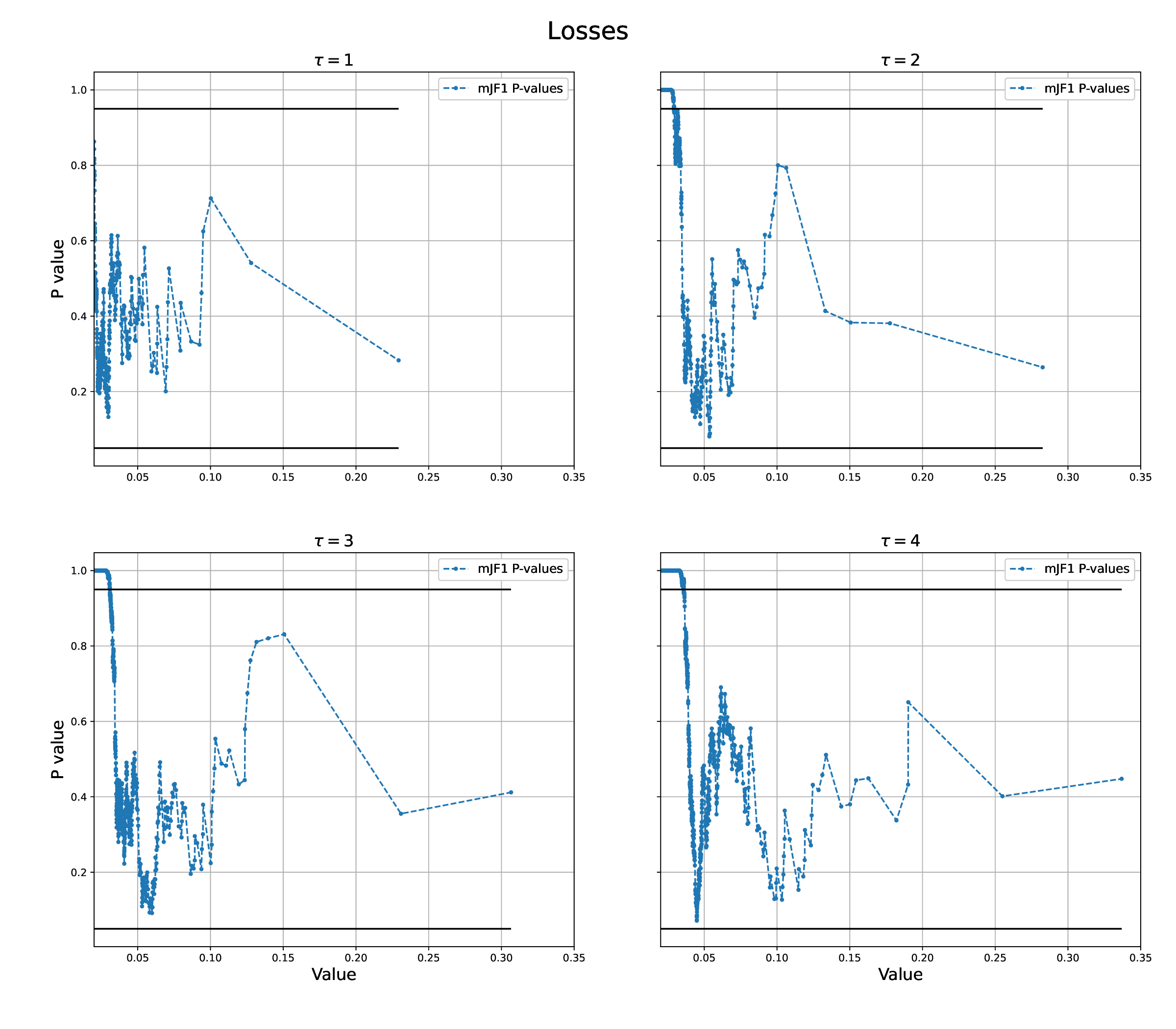}
    \caption{mJF1 p-values}
    \label{lp1-4}
\end{figure}

\begin{figure}[!htb]
    \centering
    \includegraphics[width=1\linewidth]{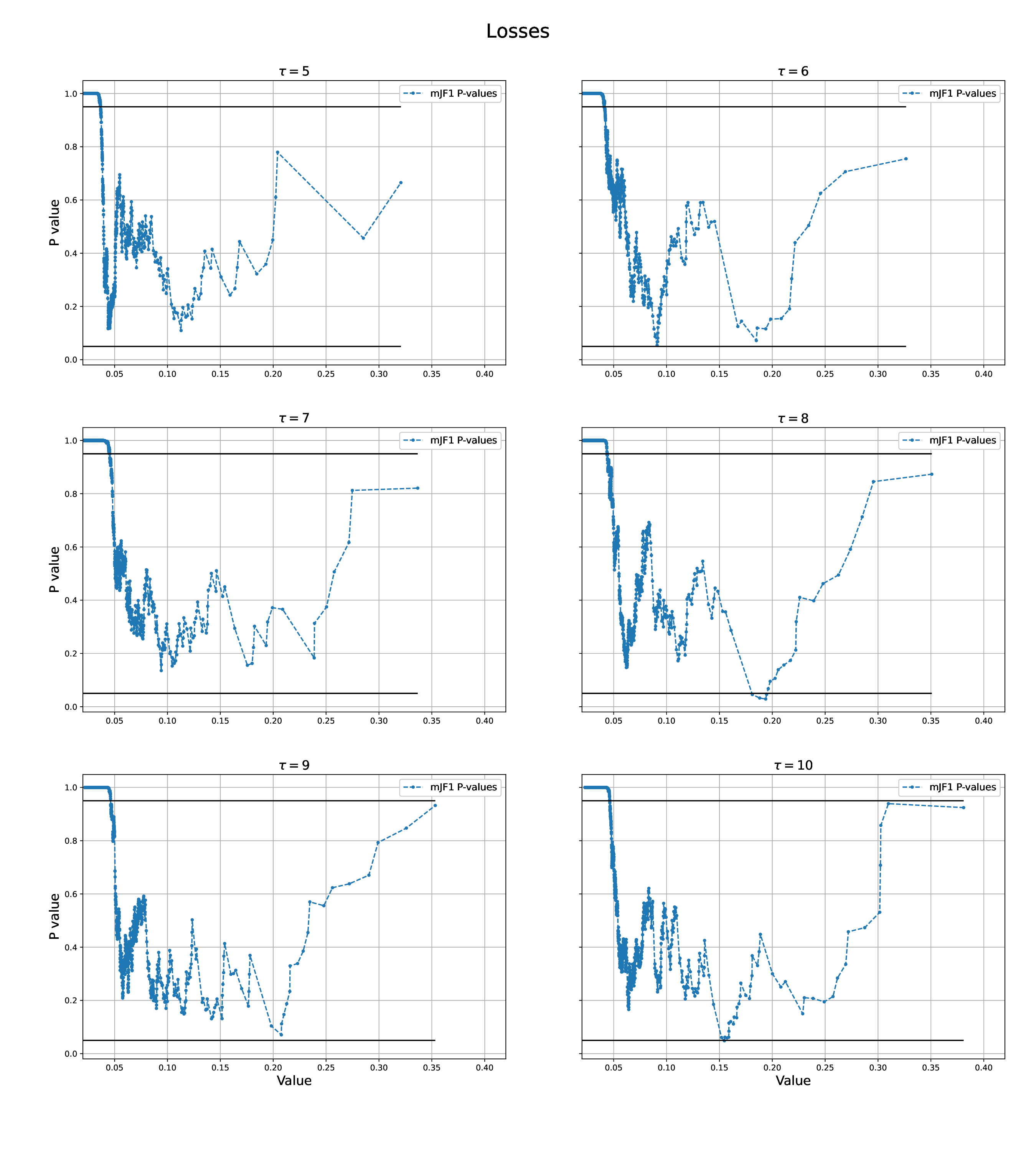}
    \caption{mJF1 p-values}
    \label{lp5-10}
\end{figure}

\end{document}